\documentclass{article}

\usepackage{amssymb}
\usepackage{latexsym}
\usepackage{booktabs}

\usepackage{url}
\usepackage{xcolor}
\definecolor{newcolor}{rgb}{.8,.349,.1}
\usepackage{multirow}
\usepackage{color}
\usepackage[title]{appendix}%
\usepackage{tikz}
\usepackage{amsmath}
\usepackage[x11names, dvipsnames]{xcolor}
\usepackage[margin=1in]{geometry}
\usepackage[colorlinks=true,citecolor=RoyalBlue,linkcolor=Magenta,urlcolor=ForestGreen]{hyperref}
\usepackage{subfig}
\hypersetup{
  colorlinks, linkcolor=red,
}

\definecolor{myblue}{HTML}{002d74}
\definecolor{myorange}{HTML}{e85113}
\definecolor{mygreen}{HTML}{1fa12e}

\usepackage{xcolor}

\usepackage{float}
\usepackage[utf8]{inputenc}  
\usepackage{subfiles} 
\usepackage[round,authoryear]{natbib}
\usepackage{amssymb}
\usepackage{mathtools}
\usepackage{gensymb}
\usepackage{authblk}

\begin{document}

\title{Bayesian Optimization for reanalysis and calibration of highly energetic sea state events simulated with a spectral third-generation wave model}

\author[1,2]{Cédric Goeury}
\author[1]{Thierry Fouquet}
\author[1]{Maria Teles}
\author[1,2]{Michel Benoit}

\affil[1]{EDF R\&D, National Laboratory for Hydraulics and Environment (LNHE), 6 Quai Watier, 78400 Chatou, France}

\affil[2]{Saint-Venant Hydraulics Laboratory, LHSV, ENPC, Institut Polytechnique de Paris, EDF R\&D, 6 Quai Watier, 78400 Chatou, France}

\date{\today}
\maketitle

\begin{abstract}
Accurate hindcasting of sea state events is a cornerstone of coastal engineering, risk assessment, and climate‑related studies, yet it remains limited by uncertainties in physical parameterizations and model structure. This study introduces an automated calibration framework based on Bayesian Optimization (BO) using the Tree‑structured Parzen Estimator (TPE) to constrain key dissipative processes in the ANEMOC‑3 hindcast wave model, including bottom‑friction losses, depth‑induced wave breaking, and dissipation driven by wave strong opposing currents. The methodology enables the joint optimization of continuous physical parameters and discrete model structure choices within a unified probabilistic search space, significantly reducing model–observation misfit. Calibration is conducted over the high energy storm conditions of February 2014, while transferability is assessed both temporally and spatially, through independent validation on January 2014 and January 2018 events and across a network of offshore and coastal buoy observations. The optimized configurations retain skill beyond the calibration period and across observation sites, yielding systematically improved agreement with buoy measurements in terms of bias, root mean square error, and scatter index relative to the reference configuration. These results highlight the potential of Bayesian Optimization as a scalable and robust framework for automating the calibration of complex wave hindcast systems. Future developments will address multi‑objective optimization, uncertainty quantification, and the integration of complementary observational datasets.
\end{abstract}

\section{Introduction}
\label{sec:intro}

Accurate hindcasting of sea state conditions plays a pivotal role in marine engineering \citep{Accensi_2021, Alday_2024}, navigational safety \citep{Rogers_2007}, and climate research \citep{Akpinar_2023}. By integrating numerical wave models with atmospheric reanalysis data, hindcasting enables the generation of long-time wave datasets that are indispensable for assessing extreme events \citep{Wurjanto_2020} and characterizing operational climates \citep{Allahdadi_2019}, particularly in regions with sparse observational coverage. These reconstructions not only inform the design and resilience of offshore and coastal infrastructure but also serve as critical benchmarks for validating forecasting systems and enhancing climate projections.

Despite significant advancements in computational capabilities and numerical modeling techniques, the reliability of hindcasted wave fields remains constrained by uncertainties inherent to model inputs, namely forcing fields \citep{Teixeira_1995, Campos_2022} and model parameters \citep{Alday_2022}, by structural assumptions, namely the modeling and parametrization choices that specify the functional form of the system \citep{Du_2019}, and by observational data \citep{Dodet_2020, Bitner_2022}. Structural and parameter uncertainties in wave models are inherently interconnected \citep{Alonso_2021}. Hindcast wave databases are typically built using third-generation wave models that solve the wave action balance equation, allowing realistic reconstruction of historical sea states. These models simulate key physical processes such as wind input, wave dissipation, and nonlinear interactions through various parameterizations embedded in source and sink terms included in the wave action balance equation. Parameterization choices define the model’s structure and critically influence its ability to represent diverse sea states and adapt to environmental variability. Differences in parameterization schemes can significantly affect wave characteristics and model fidelity \citep{Bi_2015, Kalourazi_2021}, highlighting the need for rigorous and context-specific selection.

Once a model structure is selected, parameter calibration becomes essential to accurately simulate and represent physical processes. This step involves tuning uncertain, physically-based parameters to align model outputs with observed data. In wave modeling, where empirical formulations and non-measurable parameters are common, calibration significantly enhances model fidelity and predictive performance \citep{Majidi_2023}. It reduces inherent uncertainties and strengthens the model’s reliability for both operational forecasting and scientific analysis. As highlighted by \citet{Oreskes_1994}, calibration is key to achieve ``empirical adequacy'', the degree to which a model replicates observed phenomena. 

Traditionally, calibration relies on manual, heuristic trial and error, where parameters are iteratively adjusted and evaluated after each simulation. While straightforward, this method is labor intensive, subjective, and often lacks a unique solution due to equifinality, where different parameter sets yield similar results \citep{Simmons_2017}. As an alternative, deterministic optimization methods, such as those used by \citet{Dubarbier_2015}, offer a more systematic and reproducible calibration process. However, they still fall short in fully addressing uncertainties in both observations and model structure \citep{Alonso_2021}.

Stochastic approaches provide robust alternatives to manual, deterministic calibration methods for parameter estimation and uncertainty quantification. The Generalized Likelihood Uncertainty Estimation (GLUE) algorithm exemplifies a pragmatic, brute-force strategy, employing Monte Carlo sampling to evaluate parameter sets against user-defined criteria without requiring a formal likelihood function \citep{Simmons_2017, Simmons_2019, Kroon_2020}. This makes GLUE suitable for models with poorly characterized error structures. However, its reliance on uniform priors and random sampling limits efficiency and statistical rigor in high-dimensional spaces \citep{Vrugt_2018}.

Bayesian inference offers a more formal framework by integrating prior knowledge with observational data via a likelihood function, yielding posterior distributions that quantify parameter uncertainty. When implemented through Markov Chain Monte Carlo (MCMC) methods, Bayesian calibration accommodates nonlinearities and multi-modal distributions \citep{Alonso_2021}. Due to computational constraints, Bayesian methods are typically applied to approximated models, including physically-based models with coarser spatial or temporal resolutions \citep{Alonso_2021} and surrogate models that emulate complex system behavior \citep{Ruessink_2006, Solari_2022}, with prior and likelihood assumptions guiding Bayesian processes such as data assimilation \citep{Deltares_2009, goeury_2024}.

Recent advances in artificial intelligence have positioned Bayesian Optimization (BO) as a powerful and efficient solution for model calibration, particularly in the context of complex and computationally expensive models. As machine learning architectures grow in complexity, hyperparameter tuning becomes increasingly demanding, prompting the development of scalable, automated optimization frameworks and advanced search algorithms capable of navigating high-dimensional parameter spaces \citep{Akiba_2019}.

Bayesian Optimization addresses these challenges by framing calibration as a global optimization problem. Through adaptive sampling, it efficiently explores non-convex objective functions without relying on gradient information, making it especially suitable for black-box models. Moreover, its ability to scale to high‑dimensional parameter spaces and adapt to heterogeneous data sources makes it especially well suited for science and engineering applications \citep{Garnett_2023}, which often involve complex physical processes and large computational costs. While prior Bayesian Optimization studies in geophysical and wave modeling applications have typically relied on Gaussian process surrogate models to calibrate parameters within fixed model structures \citep{Mouris_2023, Tanim_2024, Mrozowska_2025}, or have been limited to simplified configurations due to computational constraints \citep{Alonso_2021, Solari_2022}, the framework introduced here enables the joint optimization of continuous physical parameters and discrete model structure choices, such as depth‑induced breaking or wave–current interaction formulations, within a unified conditional search space. This is made possible through the Tree‑structured Parzen Estimator (TPE), which naturally handles hierarchical mixed discrete and continuous hyperparameters and scales effectively to high‑dimensional, non‑Gaussian landscapes \citep{La_2025, La_2025b}. Leveraging HPC‑parallel sampling, the workflow operates directly on a full‑scale third‑generation spectral wave model without surrogate approximations and is specifically tailored for calibrating sea state hindcasts dominated by nonlinear dissipation processes, offering a scalable and operationally applicable alternative to existing Bayesian tuning methods.

The structure of this paper is as follows. Section \ref{sec:study_case} introduces the ANEMOC-3 hindcast wave numerical database and outlines the dataset employed in this study. Section \ref{sec:Gov_eq} presents the governing equations used in the simulations, along with the relevant physical and numerical parameters. Section \ref{sec:bayes_opt} describes the probabilistic framework adopted for BO, including the materials and methods applied. The results of the simulations and optimization procedures are reported in Section \ref{sec:results}, followed by a critical discussion in Section \ref{sec:discussions}. Finally, Section \ref{sec:conclusions} summarizes the main contributions of the work and offers perspectives for future research.

\section{Case Study}
\label{sec:study_case}

\subsection{ANEMOC-3 hindcast wave numerical database}

The ANEMOC hindcast database (in French \textit{``Atlas Numérique d’Etats de Mer Océaniques et Côtiers''}) is derived from numerical simulations with TOMAWAC spectral wave model \citep{Benoit_1996}, from the openTELEMAC numerical platform (\url{www.opentelemac.org}), and provides sea state conditions across various geographic domains. Developed jointly by EDF R\&D LNHE and Cerema (which stands for Centre for Studies and Expertise on Risks, the Environment, Mobility and Urban Planning) over $16$ years ago, its primary focus is on the French coastal areas of the Atlantic Ocean, English Channel, and North Sea \citep{Benoit_2008}.

\begin{figure*}[!h]
\centering
\begin{tabular}{cc}
\subfloat[Oceanic domain\label{fig:anemoc-3_oc_mesh}]{%
\includegraphics[width=0.4\textwidth]{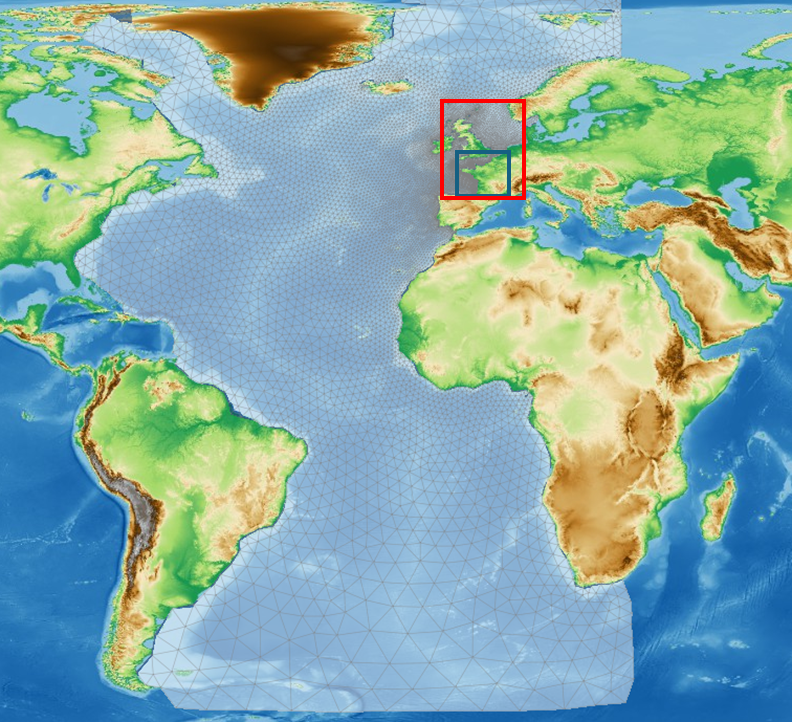}}&
\subfloat[North sea domain\label{fig:north_sea_mesh}]{%
\includegraphics[width=0.39725\textwidth]{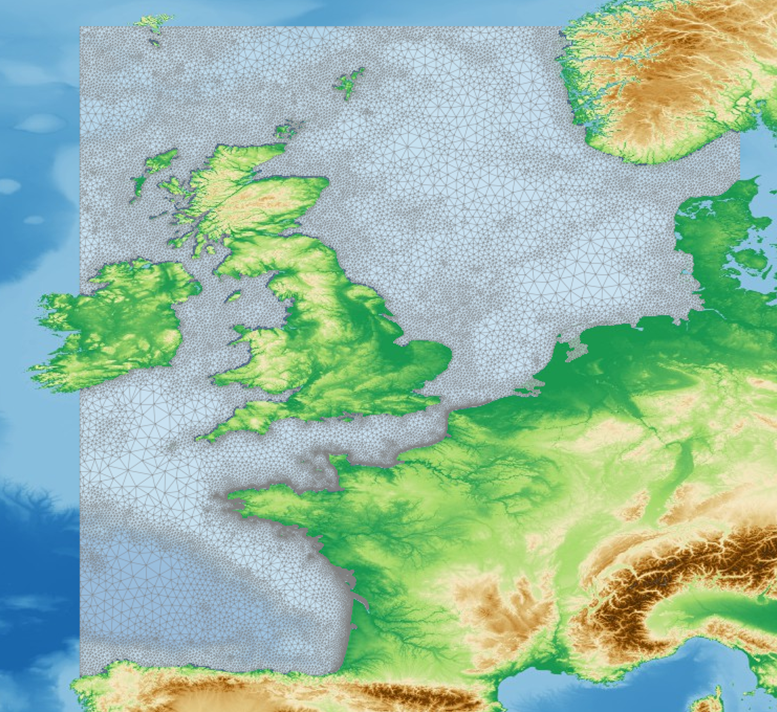}}\\
\end{tabular}
\subfloat[Coastal domain\label{fig:anemoc-3_cot_mesh}]{%
\includegraphics[width=0.43\textwidth]{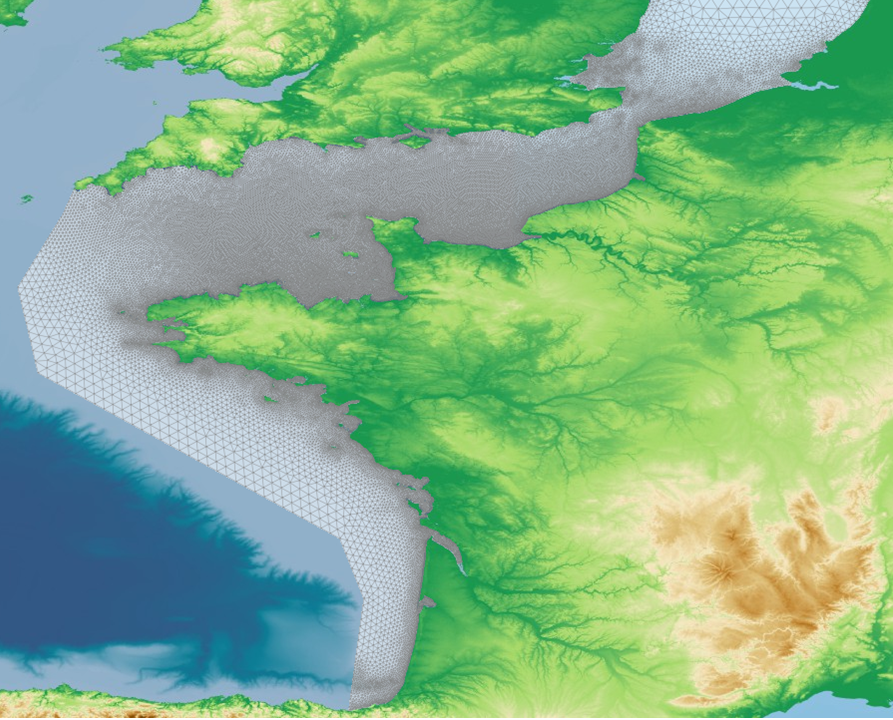}}
\caption{ANEMOC-3: Oceanic, North Sea and Coastal domains (1979$-$2024).}
\label{fig:anemoc-3_domains}
\end{figure*}

This study presents the ANEMOC‑3 version, covering the period 1979–2024, and built using two nested TOMAWAC meshes (oceanic and coastal), while a separate TELEMAC‑2D grid covering the North Sea and French coastal waters provides tidal elevations and currents to the coastal domain (Figure \ref{fig:anemoc-3_domains}). 

The oceanic mesh, comprising $12,340$  nodes and $23,542$ elements (Figure \ref{fig:anemoc-3_oc_mesh}), covers nearly the whole Atlantic ocean from 60°S to 80°N with an unstructured mesh of variable resolution, with the finest near the European waters with 0.15° resolution. The latter supplies directional variance spectra at open boundaries to the coastal mesh which has $56,375$ nodes and $108,551$ elements (Figure \ref{fig:anemoc-3_cot_mesh}), covering the French coastline beyond the $100$ m isobath with a mesh resolution up to 0.01$^\circ$, e.g roughly $1$ km. Bathymetry was interpolated from the 800 m resolution LEGOS “Europe” dataset at mean sea level, with GEBCO (General Bathymetric Chart of Oceans) data (30 arc-second resolution) supplementing uncovered regions of the oceanic mesh. Atmospheric forcing is taken from the NOAA reanalysis CFSR ($0.313^\circ \times 0.312^\circ$) and CFSv2 ($0.205^\circ \times 0.204^\circ$) data for 1979-2010 and 2011-2024 time periods respectively.

To incorporate tidal effects on wave propagation in ANEMOC-3’s coastal domain, a coupled modeling approach was implemented by \citet{Raoult_2018}, combining the TELEMAC-2D hydrodynamic model \citep{Hervouet_2007} with the TOMAWAC spectral wave model, both components of the openTELEMAC hydro-informatic system. TELEMAC-2D first operates on a mesh covering the North Sea and french coastal waters using harmonic constituents imposed at open boundaries (Figure \ref{fig:north_sea_mesh}). These are derived from the regional North-East Atlantic (NEA) tidal atlas, which provides amplitudes and phases for tidal elevation and horizontal current components \citep{Pairaud_2008,Pairaud_2010}. This first level domain gives the tidal water levels and currents to the coastal mesh, where TELEMAC-2D also operates by computing tidal water levels and currents every 15 minutes. The resulting tidal fields are taken into account by TOMAWAC to consider tidal modulation of wave dynamics on the coastal domain.

TOMAWAC solves the wave action balance equation \citep{Benoit_1996}. For the ANEMOC-3 numerical wave database, it employs a spectral discretization of 32 frequency components ranging from 0.0345 Hz to 0.6622 Hz (corresponding to wave periods between 1.51 s and 29 s), and 36 directional bins with an angular resolution of 10°. The model uses a time step of 5 min in the oceanic domain and 30 s in the coastal domain to ensure temporal resolution appropriate to each scale. The wave model explicitly accounts for the fundamental physical processes governing wave dynamics, including wind-wave generation, wave breaking dissipation, nonlinear wave interactions, and bottom friction dissipation. These processes are represented using the BAJ formulation \citep{Bidlot_2007}, which integrates Janssen’s wind input parameterization \citep{Janssen_1991}, whitecapping dissipation following  \citet{Komen_1984}, nonlinear quadruplet interactions computed via the Discrete Interaction Approximation (DIA) method \citep{Hasselmann1985}. In the coastal domain, additional mechanisms are included to account for depth-induced wave breaking, strong opposing current induced dissipation and bottom friction modeled according to the parameterization proposed by \citet{Bouws_1981}. The detailed parameterization of these processes are presented in the following sections. Further information on the configuration of ANEMOC-3 is available in \citet{Teles_2022}.

\subsection{Measurement data}
\label{subsec:observation}

\begin{figure*}[!h]
\centering
\includegraphics[width=0.75\textwidth]{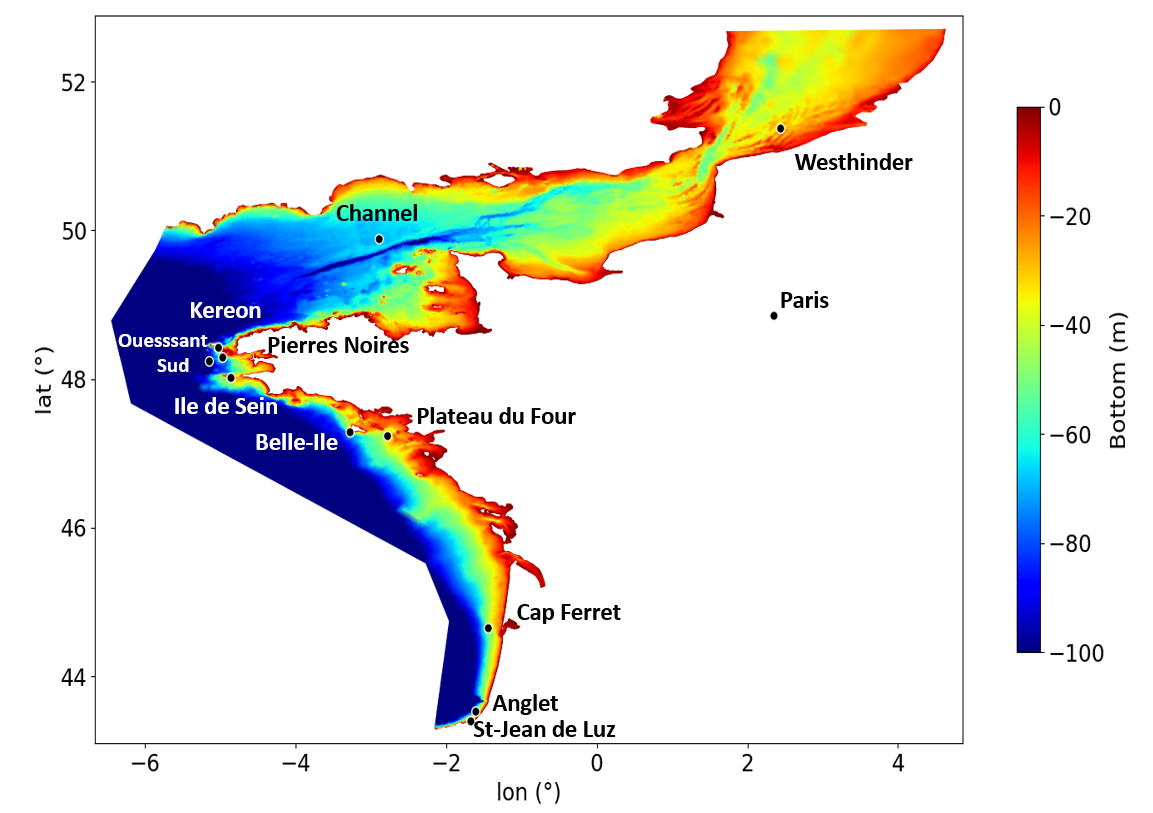}
\caption{Bottom elevation and location of buoys considered over the ANEMOC-3 coastal model.}
\label{fig:buoys_anemoc}
\end{figure*}

Wave measurements used for model calibration and validation are sourced from both offshore (deep water) and coastal (intermediate water) buoy stations, spanning water depths from approximately $20$ meters to $4300$ meters (Figure \ref{fig:buoys_anemoc}). These data are provided by the French CANDHIS network (Centre d’Archivage National des données de Houles In-Situ, \url{https://candhis.cerema.fr}), managed by Cerema, the UK Met Office (UKMO) through the WaveNet system operated by Cefas (\url{https://wavenet.cefas.co.uk}) and the \textit{Westhinder} directional wave rider buoy from the Flemish Institute for the Sea (VLIZ). The CANDHIS network comprises over $40$ waverider buoys deployed across mainland France and overseas territories, offering high-resolution wave data in real time and delayed mode. Similarly, the UKMO WaveNet network includes strategically placed buoys around the UK coastline, delivering continuous wave observations for operational forecasting and coastal monitoring.

\begin{figure*}[!h]
\centering
\includegraphics[width=0.95\textwidth]{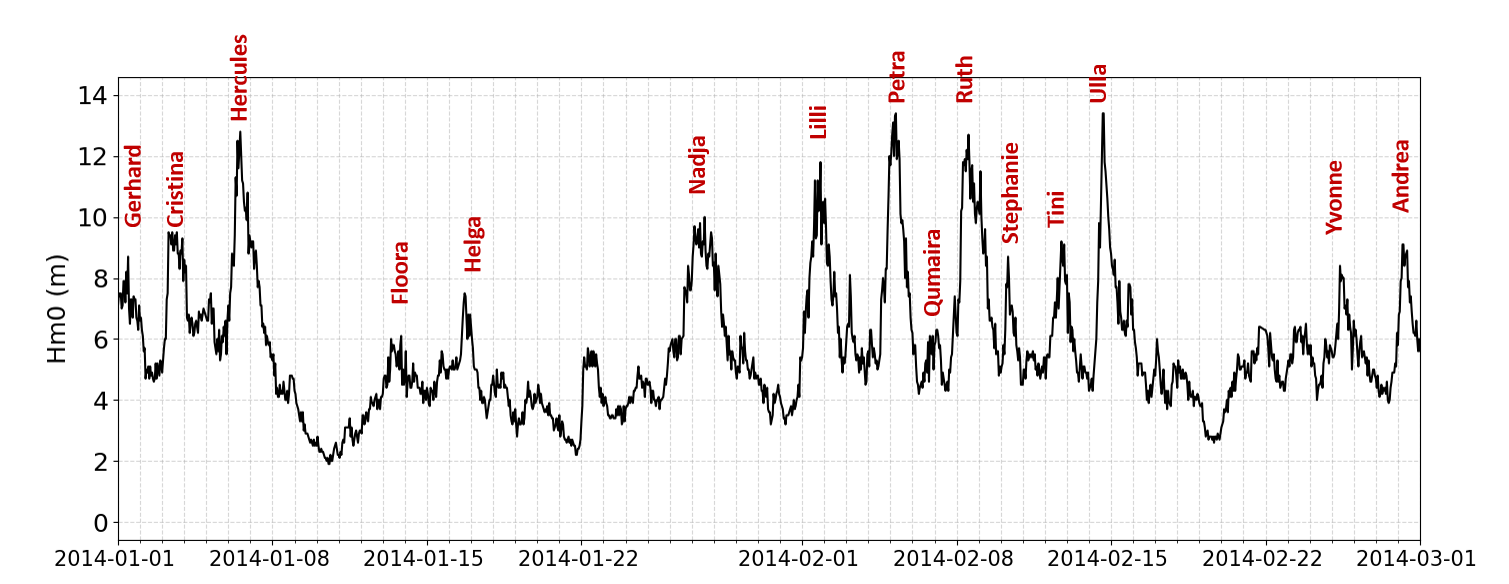}
\caption{Significant wave height evolution during the months of January and February 2014 measured at Brittany buoy (UKMO 62163) with storm names highlighted in red.}
\label{fig:brittany_fev_2014}
\end{figure*}

Spectral significant wave height ($H_{m0}$) is recorded every $30$ minutes by CANDHIS and VLIZ \textit{Westhinder} buoys and every $60$ minutes by UKMO buoys. For calibration purposes, particular focus is placed on the period of February $2014$, which was marked by a serie of intense storm events that impacted severely the Atlantic french coast. For validation, the period of February was extended to January $2014$, which was also marked by energetic wave events, notably the Hercules storm. As illustrated in Figure \ref{fig:brittany_fev_2014}, the Brittany buoy (UKMO $62163$), positioned offshore Brittany region in the Atlantic Ocean, recorded five storms during the January-February period, each with maximum values of significant wave heights ($H_{m0}$) exceeding 10 meters. Notably, during the first half of February $2014$, $H_{m0}$ remained consistently above $4$ meters, with maximum values reaching up to $13$–$14$ meters. This high-energy period provides a robust dataset for calibrating and validating wave model performance under such extreme conditions. In addition, the period of January $2018$ was also retained for independent cross-validation analyses, as detailed in Section~\ref{subsec:cross_validation_global}.

\section{Governing equations}
\label{sec:Gov_eq}

\subsection{The wave action conservation equation}

TOMAWAC numerically solves the wave action balance equation, including energy source and dissipation mechanisms through semi-empirical parameterizations. The principal variable used to characterize the sea state is the directional wave energy spectrum, hereafter denoted by \( E \). This spectrum provides a comprehensive representation of the distribution of wave energy across frequencies and directions. From the directional spectrum \( E \), the wave action density \( N \) can be derived using the following relationship:

\begin{equation}
    N = E/\rho g \sigma = F/\sigma 
\end{equation}

where $\sigma$ denotes the relative or intrinsic angular frequency, i.e. the angular frequency being observed in a coordinate system moving at the velocity of current, $\rho$ the water density, $g$ the gravitational acceleration and $F$ the directional variance spectrum.

TOMAWAC solves the following action flux conservation or balance equation:

\begin{equation}
\label{eq:balance_eq}
    \frac{\partial N}{\partial t} + \frac{\partial (\dot{x} N)}{\partial x} + \frac{\partial (\dot{y} N)}{\partial y} + \frac{\partial (\dot{k_x} N)}{\partial k_x} + \frac{\partial (\dot{k_y} N)}{\partial k_y} = Q(k_x, k_y, x, y, t)
\end{equation}

where $x$ and $y$ are the horizontal Cartesian coordinates, $t$ is time, $k_x$ and $k_y$ the wave number for directional spectrum discretization along $x$ and $y$ coordinates respectively and $Q(k_x,k_y,x,y,t)$ are the source, transfer and sink terms. The dot notation (e.g., \( \dot{x} \), \( \dot{k_x} \)) denotes the rates of change of the respective variables, reflecting wave propagation in physical and spectral space.

The source and sink terms that compose $Q(k_x,k_y,x,y,t)$ in the right-hand members of Eq.~\ref{eq:balance_eq} gather the contributions from the physical processes listed in Eq.~\ref{eq:term_source}.

\begin{align}
\label{eq:term_source}
Q =\ &
\underbrace{\begin{array}{c}
\textcolor{blue}{Q_{\text{in}}} \\
\textcolor{black}{{\scriptsize \text{Wind}}}\\
\textcolor{black}{{\scriptsize \text{input}}}
\end{array}}_{\textcolor{blue}{\text{Generation}}}+\nonumber \\[1ex]
& \underbrace{\begin{array}{c}
\textcolor{orange}{Q_{\text{ds}}} \\
\textcolor{black}{{\scriptsize\text{Whitecapping}}}\\
\textcolor{orange}{{\scriptsize\text{}}}
\end{array}
+ 
\begin{array}{c}
\textcolor{orange}{Q_{\text{bf}}} \\
\textcolor{black}{{\scriptsize\text{Bottom}}}\\
\textcolor{black}{{\scriptsize\text{friction}}}
\end{array}
+ 
\begin{array}{c}
\textcolor{orange}{Q_{\text{br}}} \\
\textcolor{black}{{\scriptsize\text{Breaking}}}\\
\textcolor{black}{{\scriptsize\text{}}}
\end{array}
+ 
\begin{array}{c}
\textcolor{orange}{Q_{\text{ds,cur}}} \\
\textcolor{black}{{\scriptsize\text{Strong}}}\\
\textcolor{black}{{\scriptsize\text{current}}}
\end{array}
+ 
\begin{array}{c}
\textcolor{orange}{Q_{\text{veg}}} \\
\textcolor{black}{{\scriptsize\text{Vegetation}}}\\
\textcolor{black}{{\scriptsize\text{}}}
\end{array}
+ 
\begin{array}{c}
\textcolor{orange}{Q_{\text{porous}}} \\
\textcolor{black}{{\scriptsize\text{Porous}}}\\
\textcolor{black}{{\scriptsize\text{medium}}}\\
\end{array}}_{\textcolor{orange}{\text{Dissipation}}}
+ \nonumber \\[1ex]
&\underbrace{\begin{array}{c}
\textcolor{green!60!black}{Q_{\text{nl}}} \\
\textcolor{black}{\scriptsize\text{\scriptsize{Quadruplets}}}
\end{array}
+ 
\begin{array}{c}
\textcolor{green!60!black}{Q_{\text{tr}}} \\
\textcolor{black}{\text{\scriptsize Triads}}
\end{array}}_{\textcolor{green!60!black}{\text{Transfers within the spectrum}}}
\end{align}

In this study, particular attention is given to dissipation mechanisms that are especially relevant in the coastal domain of ANEMOC-3, characterized by limited-depth environments. These include the bottom friction dissipation term (\( Q_{\text{bf}} \)), the depth-induced breaking dissipation term (\( Q_{\text{br}} \)), and the wave dissipation associated with strong opposing currents (\( Q_{\text{ds,cur}} \)). These processes are critical in governing wave energy attenuation and transformation in shallow water regions, where their influence on the resulting sea state becomes increasingly significant. A detailed description of each term is provided in the following sections.

\subsection{Sink terms and parameters}
\label{sec:wave parameters}
\subsubsection{Bottom friction-induced dissipation}

Bottom friction-induced wave energy dissipation is modeled using an empirical formulation that captures the combined effects of wave–seabed interactions, including percolation and frictional resistance. In this study, a linearized version of the original expression proposed by \citet{Hasselmann_1973} is employed, which simplifies the representation while retaining the essential physical characteristics of the dissipation process. The adopted formulation is expressed as follows:

\begin{equation} 
\label{eq:bot_fric} 
Q_{\text{bf}} = -\Gamma \frac{2k}{g \sinh(2kh)} F
\end{equation}

where $\Gamma$ is a constant dissipation coefficient, $h$ is the local water depth, and $k=\sqrt{k_x^2 + k_y^2}$ is the wave number magnitude. 

Empirical studies have provided a range of values for the bottom friction dissipation coefficient, $\Gamma$, depending on seabed characteristics and wave conditions. In the JONSWAP experiment, \citet{Hasselmann_1973} proposed a value of $\Gamma = 0.038 \, \mathrm{m}^2 \, \mathrm{s}^{-3}$ for swell dissipation over sandy bottoms. This was later revised by \citet{Bouws_1981}, who suggested $\Gamma = 0.067 \, \mathrm{m}^2 \, \mathrm{s}^{-3}$ for fully developed wind seas. Subsequent research has broadened the applicable range to values between 0.003 and 0.15 \citep{Padilla_2001, Cialone_2007, Bastidas_2016}. These findings collectively support the adoption, in the current study, of a generalized range of $0.01$ to $0.1$ for $\Gamma$ in wave models, providing a balance between physical realism and model flexibility across diverse seabed conditions.

\subsubsection{Depth-induced breaking dissipation}

Depth-induced wave breaking is represented through parametric spectral formulations that quantify the associated energy dissipation while preserving the shape of the directional spectrum. Originally developed for random sea states, these parameterizations rely on an analogy with hydraulic jumps, whereby breaking acts as an energy sink without redistributing energy across frequencies or directions. In the present study, the dissipation rate is evaluated using a characteristic frequency selected as either the spectral peak or an appropriate mean frequency. Two formulations are considered in the current study. The first, following \citet{Battjes_1978}, assumes that all breaking waves are limited by a maximum attainable height determined by the local water depth or by a steepness criterion. The model truncates a Rayleigh wave-height distribution at this limiting height, from which it computes the fraction of breaking waves. The resulting dissipation is then applied uniformly across the frequency–direction spectrum. The second approach, based on \citet{Thornton_1983}, represents dissipation through statistical descriptions of breaking wave heights. Two formulations are commonly used. The first assumes a uniform distribution in which all waves contribute equally to the energy loss. The second relies on a weighted distribution that assigns greater importance to larger waves, in accordance with field observations. In both cases, the resulting energy sink depends on empirical breaking parameters and the local water depth, while remaining uniform across the spectrum.

The detailed mathematical expressions of the depth-induced breaking dissipation source terms for both parameterizations, together with all auxiliary definitions, are provided in \ref{app:breaking}. The following paragraphs detail the empirical parameters governing each depth-induced breaking formulation as implemented in TOMAWAC, together with the calibration ranges adopted in this study.

\begin{itemize}
    \item Battjes and Janssen’s model
\end{itemize}

In the TOMAWAC implementation of the \citet{Battjes_1978_conf} depth-induced breaking formulation, three empirical coefficients control the dissipation rate: $\gamma_1$, $\gamma_2$, and $\alpha$. The model defaults are $\gamma_1 = 0.88$, $\gamma_2 = 0.80$, and $\alpha = 1.0$. Physically, $\gamma_2$ sets the maximum individual wave height relative to local depth, defining the onset of depth-limited breaking; $\gamma_1$ adjusts the shape of the breaking probability function and the effective breaker height, influencing the fraction of waves assumed to break; and $\alpha$ is a linear scaling factor on the dissipation term, directly controlling the magnitude of energy loss once breaking occurs. Because the \citet{Battjes_1978_conf} formulation relies on simplified assumptions, such as Rayleigh statistics for wave heights and a hydraulic jump analogy (i.e., treating wave breaking as similar to a sudden water level rise for estimating energy dissipation), the associated coefficients do not have universal values and must be calibrated to local conditions.

Field and laboratory calibrations consistently report breaker indices deviating from the original value of $\gamma_2 = 0.80$, with observed values ranging between approximately 0.4 and 1.5 depending on bathymetry, slope, and spectral shape \citep{Battjes_1985, kraus1994, Apotsos_2008, carini2021}. Similarly, $\alpha$ has been calibrated in the literature over a range of approximately 0.2 to 1.2, and occasionally higher when compensating for other source-term settings \citep{Rattanapitikon_2002, Damlamian_2013}. Based on these findings, conservative calibration envelopes of $\gamma_1 \in [0.50, 1.20]$, $\gamma_2 \in [0.50, 1.20]$, and $\alpha \in [0.50, 1.50]$ are considered for this study. These ranges bracket theoretical benchmarks, model defaults, and the spread of empirical fits, while ensuring that site-specific optimal values can be identified through the calibration process.

\begin{itemize}
    \item Thornton and Guza’s model
\end{itemize}

The depth-induced wave breaking formulation developed by \citet{Thornton_1983} relies on two empirical parameters: $\gamma$, which modulates the spectral distribution of wave energy dissipation, and $b$, which scales its overall intensity. In TOMAWAC, the default parameter values are set to $\gamma = 0.42$ and $b = 1.0$; nevertheless, the model documentation highlights the critical need for site-specific calibration in order to accurately represent local wave transformation processes and energy dissipation dynamics. Allowing $\gamma$ to vary within the interval $[0.2, 0.6]$ is consistent with both laboratory and field studies, where lower values tend to distribute dissipation more broadly across the wave spectrum, while higher values concentrate it on dominant wave components \citep{Thornton_1983, Apotsos_2008, Gon_2019}. Similarly, the range $[0.5, 2.0]$ for $b$ is supported by calibration exercises in coastal wave modeling, where $b$ serves as a key tuning parameter to match observed wave decay and surf-zone setup \citep{Thornton_1983, Westhuysen_2010}. These intervals offer sufficient flexibility to accommodate site-specific variability while maintaining physically realistic dissipation behavior.

\subsubsection{Wave blocking effects}

Wave blocking arises when waves propagate against an opposing current whose magnitude approaches the wave group velocity, ultimately reducing or suppressing energy transmission. Accurate representation of this process is essential in regions with strong ambient currents. In spectral wave models, blocking is commonly represented using either (i) an equilibrium limitation of the spectrum, based on a Phillips-type upper bound that constrains the high-frequency tail under strong opposing currents, or (ii) an additional dissipative source term that enhances whitecapping dissipation as current induced steepening increases. The latter approach, as proposed by \citet{Westhuysen_2012}, provides an explicit and physically consistent representation of current induced wave breaking through a modification of the dissipation term.

The \citet{Westhuysen_2012} formulation accounts for the influence of current velocity on wave breaking by linking enhanced dissipation to the intrinsic frequency shift, the local level of spectral saturation, and empirical tuning parameters. Two coefficients govern this mechanism: the saturation threshold \( B_r \), which controls the onset of nonlinear wave breaking, and the scaling factor \( C_{ds,\text{cur}} \), which adjusts the dissipation intensity in response to the opposing current. In the present study, \( C_{ds,\text{cur}} \) is varied within the range \( [0.4, 0.9] \), following the original recommendations. The saturation threshold \( B_r \) plays a key role in determining the initiation of breaking. Laboratory and field observations indicate that wave breaking typically occurs when saturation exceeds values between 0.001 and 0.002 \citep{Banner_2002}. This range was adopted and validated in the revised whitecapping dissipation formulation by \citet{Westhuysen_2007}, who showed that lower saturation thresholds improve model performance across both deep and shallow water environments. These parameter ranges ensure consistency with observed wave-current interactions and support the physical realism of the dissipation behavior.

The complete mathematical expressions and auxiliary definitions associated with these two complementary approaches for representing wave–current interaction are provided in ~\ref{app:blocking}.

\section{Bayesian Optimization for Model Tuning}
\label{sec:bayes_opt}

Section \ref{sec:Gov_eq} introduced multiple parameterizations and calibration ranges for key wave dissipation processes. The present section describes the Bayesian Optimization framework employed to explore this parameter space and identify the model configurations that best reproduce observed sea states. The exposition below follows the treatment of Bayesian Optimization (BO) presented in \citet{Garnett_2023}, to which the interested reader is referred for a comprehensive introduction and additional theoretical insights.

BO is a strategy for efficiently minimizing a black-box objective function \( f : \mathcal{X} \to \mathbb{R} \), where \(\mathcal{X}\) denotes the space of feasible parameter configurations. The aim is to identify a global minimizer
\begin{equation}
\boldsymbol{\theta}_{\mathrm{opt}} \in \arg\min_{\boldsymbol{\theta} \in \mathcal{X}} f(\boldsymbol{\theta}),
\label{eq:minization}
\end{equation}
while limiting costly evaluations and without requiring an analytic expression for \(f\). 

Consider a dataset \( \mathcal{D}_n = \{(\boldsymbol{\theta}_i, \phi_i)\}_{i=1}^n \), where each objective function evaluation is explicitly assumed to be noise free, so that $\phi_i = f(\boldsymbol{\theta}_i)$. This noise free assumption constitutes a deliberate modeling choice made in the present work. Under this framework, BO formulates the continued  search for the optimal parameter vector $\boldsymbol{\theta}_{\mathrm{opt}}$ as a sequential decision-making problem. The optimizer must select the next evaluation points based solely on the accumulated information in $\mathcal{D}_n$. To support these decisions, BO provides a principled framework for reasoning under uncertainty. Each evaluation choice is modeled as a decision whose action space is the domain $\mathcal{X}$, and because the true objective $\phi=f(\boldsymbol{\theta})$ is unknown, these actions are taken under uncertainty regarding their outcomes. BO therefore maintains a probabilistic belief \( p(\phi \mid \boldsymbol{\theta}, \mathcal{D}_n) \) over the unknown objective function value \( \phi \), given the set of previously evaluated parameter configurations \( \mathcal{D}_n \), and updates this belief as new evaluations are performed. This posterior predictive distribution allows the optimizer to anticipate the consequences of evaluating the objective $\phi$ at any candidate point \( \boldsymbol{\theta} \), making it a central ingredient of the search.

Determining which evaluations are most valuable requires specifying what constitutes a desirable outcome of the optimization process. To this end, BO introduces a utility function $u(\mathcal{D}_n)$ that encodes preferences over the datasets that could be returned at the end of the search. By selecting actions that, in expectation, yield the greatest improvement in this utility, BO defines a coherent and adaptive strategy for sequentially acquiring information that is most informative for identifying the global minimum of the objective function. However, directly selecting the next evaluation point by maximizing expected utility is generally infeasible, as it would require integrating over the full posterior of the unknown function and all possible future datasets, a task that quickly becomes computationally prohibitive. Bayesian Optimization therefore relies on an acquisition function, which provides a tractable criterion for selecting the next evaluation based on the current posterior information.

\subsection{The expected improvement acquisition function}

Expected Improvement (EI) is one of the most widely adopted acquisition policies in Bayesian Optimization. Consider the problem of identifying the point in the domain that minimizes the objective function, with the ultimate goal of recommending one of the evaluated locations for long-term use. Conditional on the current dataset of evaluations $\mathcal{D}_n$, the EI criterion quantifies the expected gain obtained by evaluating the objective at a candidate location $\boldsymbol{\theta}$ and augmenting $\mathcal{D}_n$ with the resulting evaluation. More precisely, EI measures the expected reduction in the best achievable value of the objective, equivalently, the expected increase in utility, that would arise from incorporating this prospective evaluation into the posterior model.

Given that the present work assumes exact (noiseless) evaluations of the objective  function, observing a location \( \boldsymbol{\theta} \) deterministically yields \( \phi = f(\boldsymbol{\theta}) \). The utility function is defined as the minimum objective value in the dataset such as $u(D_n) = \min_{i \in \{1,\dots,n\}} \phi_i=\phi^\ast$. A new sample \((\boldsymbol{\theta}, \phi)\) therefore contributes utility only if it improves upon the current minimum, resulting in a marginal gain $u(D_n\cup \{(\boldsymbol{\theta},\phi)\}) - u(D_n) = \max(\phi^\ast - \phi,\, 0)$. Taking the expectation of this quantity with respect to the predictive distribution leads directly to the classical Expected Improvement (EI) acquisition function which quantifies the expected increase in the best-observed value when querying a new point $\boldsymbol{\theta}$:

\begin{equation}
\alpha_{\mathrm{EI}}(\boldsymbol{\theta}, D_n)
\;=\;
\int \max\!\bigl(\phi^\ast - \phi,\, 0\bigr)\; p(\phi \mid \boldsymbol{\theta}, D_n)\, d\phi.
\label{eq:expected_improvement}
\end{equation}

As noted by \citet{Garnett_2023}, Expected Improvement naturally diminishes in regions that have already been explored, as further evaluations there are unlikely to yield substantial gains. By favoring points that combine low predicted values with high uncertainty, EI therefore balances exploitation of promising regions with exploration of less sampled areas.

Estimating the probability \( p(\phi\mid\boldsymbol{\theta},\mathcal{D}_n) \) has been widely studied, with various methods proposed in the literature. Gaussian Processes (GPs) are widely used  surrogate models in BO due to their analytically tractable posterior distributions \citep{Song_2022}. In this work, the Tree-structured Parzen Estimator (TPE) \citep{Bergstra_2011} is adopted for BO due to its practical advantages over Gaussian Processes (GPs). By modeling the inverse conditional density \( p(\boldsymbol{\theta}\mid \phi, \mathcal{D}_n) \) and applies Bayes' theorem to recover the posterior, TPE offers enhanced \textit{flexibility}, by natively supporting categorical and conditional hyperparameters through its tree-structured search space \citep{Watanabe_2023}, a setting in which GP‑based models often require bespoke kernel engineering or relaxations that degrade performance. Moreover, TPE demonstrates superior \textit{scalability} \citep{Bergstra_2011}, avoiding the \(\mathcal{O}(n^3)\) complexity of GP inference caused by covariance matrix inversion; it also exhibits more stable behavior as the number of evaluations increases, since its non‑parametric density estimators scale more gracefully than GP kernels. Finally, TPE provides greater \textit{efficiency} \citep{Ozaki_2022} by leveraging non-parametric density estimation to capture multi-modal, non-Gaussian distributions and enable parallel, asynchronous evaluations, a capability that is more challenging to achieve reliably with GP surrogates.

\subsection{Tree-structured Parzen Estimator (TPE)}

\citet{Bergstra_2011} showed that the Expected Improvement (EI) acquisition function can be reformulated as a density ratio estimation problem. Given a threshold $\phi^\ast$, typically defined as a quantile $q \in (0,1)$ of the observed objective values such that $p(\phi \le \phi^\ast) = q$, two conditional probability density functions are introduced:

\begin{equation}
\left\{
\begin{array}{l}
l(\boldsymbol{\theta}) = p\left(\boldsymbol{\theta}\mid\phi\le\phi^\ast,\mathcal{D}_n\right)\\
g(\boldsymbol{\theta})=p\left(\boldsymbol{\theta}\mid\phi>\phi^\ast,\mathcal{D}_n\right)
\end{array}
\right.
\end{equation}

where, $\ell(\boldsymbol{\theta})$ is constructed from the evaluated parameter configurations $\{\boldsymbol{\theta}_i\}$ whose corresponding objective values $\phi_i = f(\boldsymbol{\theta}_i)$ satisfy $\phi_i \le \phi^\ast$, while $g(\boldsymbol{\theta})$ is constructed from the remaining evaluated configurations.

The densities $\ell(\boldsymbol{\theta})$ and $g(\boldsymbol{\theta})$ are estimated using one‑dimensional Parzen kernel density estimators applied independently to each active component of $\boldsymbol{\theta}$. Categorical choices define a tree‑structured search space, determining which parameters are active within each branch. A key result of \citet{Bergstra_2011} is that the EI acquisition function, $\alpha_{\mathrm{EI}}(\boldsymbol{\theta},\mathcal{D}_n)$, can be expressed in terms of the density ratio $\ell(\boldsymbol{\theta})/g(\boldsymbol{\theta})$. Maximizing EI is therefore equivalent to maximizing this ratio, favoring parameter configurations that resemble previously well‑performing ones.

In practice, the TPE algorithm proceeds by: (i) selecting a quantile threshold $\phi^\ast$; (ii) partitioning past evaluations into two subsets, $\mathcal{D}^\ell = \{(\boldsymbol{\theta}_i,\phi_i): \phi_i \le \phi^\ast\}$ and $\mathcal{D}^g = \{(\boldsymbol{\theta}_i,\phi_i): \phi_i > \phi^\ast\}$; (iii) estimating $\ell(\boldsymbol{\theta})$ and $g(\boldsymbol{\theta})$ using Parzen estimators; and (iv) generating candidate configurations by sampling from $\ell(\boldsymbol{\theta})$ and ranking them according to the ratio $\ell(\boldsymbol{\theta})/g(\boldsymbol{\theta})$. In the batch setting considered in this study, this sampling procedure is repeated multiple times to generate a set of candidate configurations before any corresponding objective values become available. In this study, $\phi^\ast$ is chosen such that the top 10\% of evaluated trials, with a maximum of 25, define $\mathcal{D}^\ell$, following the original recommendations of \citet{Bergstra_2011}. This choice provides an effective balance between exploration and exploitation while maintaining stable density estimates. The open-source hyperparameter optimization framework Optuna \citep{Akiba_2019} is used to implement TPE and manage the optimization workflow. For a detailed discussion of the TPE methodology and its components, the reader is referred to \citet{Watanabe_2023}.

\subsection{Practical strategies for physical link implementation}
\label{sub:practial_part}

As presented in Section \ref{sec:Gov_eq}, the wave dynamics model employs empirical parameterizations to represent essential physical processes, including bottom friction, depth-induced wave breaking, and wave blocking due to opposing currents, which are the focus of this study. The control vector $\boldsymbol{\theta}$, introduced in Eq.~\ref{eq:minization}, is composed of the corresponding model parameters. Since TOMAWAC (thereafter denoted $M$) operates on a spatially and temporally discrete framework, each realization $\boldsymbol{\theta}_j$ produces a model output $Y$ that corresponds to buoy observations (see Subsection \ref{subsec:observation}). For each point of interest ${\bf{x}}_{p}$, representing buoy-equivalent locations, the output consists of significant wave height values at discrete time steps $t \in [1, \dots, T]$, expressed as $Y = (Y_1, \dots, Y_T)$, where $Y_i = Y({\bf{x}}_{p}, t_i) = M(\boldsymbol{\theta}_j; ({\bf{x}}_{p}, t_i))$. Reliable wave prediction across coastal environments requires that the numerical model closely aligns with historical observations of significant wave height. This alignment is achieved through a calibration process that minimizes discrepancies between model outputs and measurements at observation stations, by optimizing the following error metric:

\begin{equation}
    \label{eq:error_metrics}
    f\left(\boldsymbol{\theta}\right) = \frac{1}{2}\left(\textbf{Y}^{obs}-\mathcal{H}\left(M(\boldsymbol{\theta})\right)\right)^T\mathcal{W}^{-1}\left(\textbf{Y}^{obs}-\mathcal{H}\left(M(\boldsymbol{\theta})\right)\right)
\end{equation}

where \( \mathcal{W} \) is a weighting matrix that adjusts the relative influence of each observation based on local depth, assigning greater weight to shallower buoy data to enhance sensitivity to depth-dependent wave behavior, $\mathcal{H}$ is an operator that projects the numerical output onto the observation space and \( \textbf{Y}^{obs} \in \mathbb{R}^{N_{\text{obs}}} \) denotes the vector of observed significant wave heights at buoy locations.

This choice is motivated by the use of the spectral significant wave height, $H_{m0}$, a robust observational parameter characterized by relatively low measurement noise. As an integral measure of total wave spectral energy, $H_{m0}$ provides a reliable representation of overall sea state conditions and is well suited for model calibration. However, reliance on an integral sea state metric introduces a degree of non‑uniqueness, as distinct combinations of model parameters or physical parameterizations may yield comparable misfit values, reflecting the issue of equifinality commonly encountered in wave model calibration. Rather than attempting to explicitly resolve this ambiguity, the Bayesian Optimization framework is designed to efficiently explore and rank competing solutions within the admissible parameter space. Consequently, the resulting optimum should be interpreted as representative of a family of near‑optimal, physically consistent configurations rather than a uniquely identifiable parameter set. At this stage, a critical question arises: which observations (\(\textbf{Y}^{obs}\)) should be incorporated into the calibration process? This issue is examined in detail in the following section.

\subsubsection{Selection of the calibration data subset}

To investigate the spatial variability of model errors across buoy measurement sites, a Principal Component Analysis (PCA) was performed on the error fields. These errors were computed by comparing buoy observations with outputs from 1,000 ANEMOC-3 simulations generated using a Monte Carlo design of experiments. The model structure incorporated depth-induced breaking following \citet{Thornton_1983} and wave dissipation due to opposing currents as formulated by \citet{Westhuysen_2012}. PCA, a dimensionality reduction technique, transforms the original variables into a new set of orthogonal (uncorrelated) principal components that capture the maximum variance in the data.

\begin{figure*}[!h]
\centering
\includegraphics[width=0.7\textwidth]{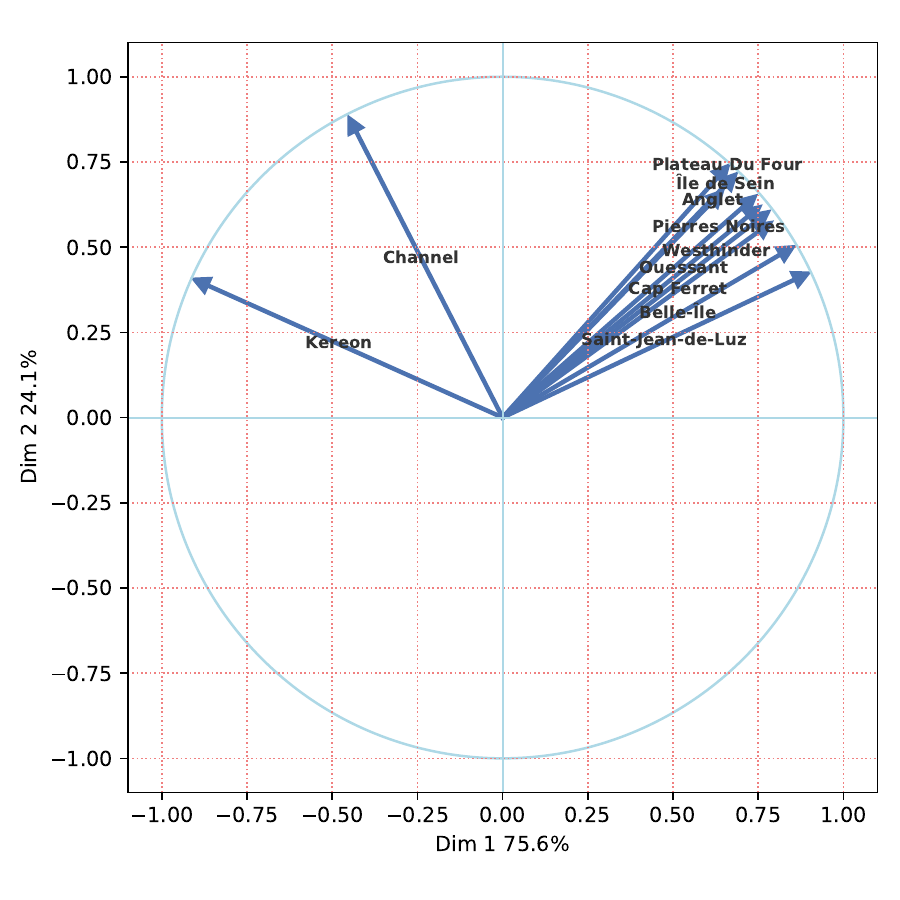}
\caption{Correlation circle of the first two principal components of the PCA (explaining more than 99\% of the total variance), illustrating model errors \(f(\boldsymbol{\theta})\) across buoy measurement sites based on 1,000 Monte Carlo simulations using a model structure incorporating depth-induced breaking \cite{Thornton_1983} and wave dissipation due to opposing currents \cite{Westhuysen_2012}.}
\label{fig:corr_circle}
\end{figure*}

The PCA correlation circle shown in Figure \ref{fig:corr_circle} illustrates the correlation between model errors at the buoy locations, projected onto the first two principal components, which together explain more than 99\% of the total variance.

\newpage

The correlation circle reveals distinct groupings among the buoy stations based on the similarity of their associated model errors. Two clearly separated clusters are observed: one comprising the \textit{Kereon} and \textit{Channel} buoys, and another including the remaining stations. To avoid introducing a trade-off between these divergent groups during the calibration process, the \textit{Kereon} and \textit{Channel} buoys were excluded.

Among the remaining buoys, the angle between vectors in the correlation circle reflects the degree of correlation between their associated errors. Based on this criterion, three groups of correlated buoys can be identified: (i) \textit{Plateau~du~Four}, \textit{Île~de~Sein}, and \textit{Anglet}; (ii) \textit{Pierres~Noires}, \textit{Ouessant}, \textit{Cap~Ferret}, and \textit{Westhinder}; and (iii) \textit{Belle--Île} and \textit{Saint--Jean--de--Luz}.

Although the error associated with the \textit{Anglet} buoy is closely aligned with that of \textit{Île~de~Sein} in the PCA space, its slightly weaker correlation and collinearity led to its exclusion in favor of the more representative \textit{Île~de~Sein}.

To maintain balanced representation across groups and avoid overemphasizing any single cluster, the group composed of \textit{Pierres~Noires}, \textit{Ouessant}, \textit{Cap~Ferret}, and \textit{Westhinder} was reduced to two buoys. In this context, the two deeper buoys (\textit{Ouessant} and \textit{Westhinder}) were excluded to enhance sensitivity to depth-dependent wave behavior.

In summary, based on the correlation analysis, the buoy stations retained for model calibration in this study are: \textit{Plateau~du~Four}, \textit{Île~de~Sein}, \textit{Pierres~Noires}, \textit{Cap~Ferret}, \textit{Belle--Île}, and \textit{Saint--Jean--de--Luz}.

\section{Results}
\label{sec:results}

The results are structured around three complementary components. First, the behavior of the Bayesian Optimization framework is analyzed, with particular emphasis on its ability to efficiently explore the high‑dimensional mixed discrete and continuous parameter space governing wave dissipation processes, including convergence properties and sampling dynamics, and to identify a stable, physically consistent optimal parameter set. Second, the physical performance of the calibrated model is assessed, focusing on improvements in the representation of high-energy sea states and on the physical interpretation of the inferred parameter adjustments in a coastal and oceanic context. Finally, the robustness and generalizability of the optimized parameters are evaluated through cross‑validation experiments conducted at independent buoy locations and over distinct storm‑active periods, thereby demonstrating the transferability of the calibrated configuration across spatial domains and winter storm regimes.

\subsection{Bayesian Optimization Dynamics and Parameter Space Exploration}

\begin{figure*}[!h]
\centering
\includegraphics[width=0.75\textwidth]{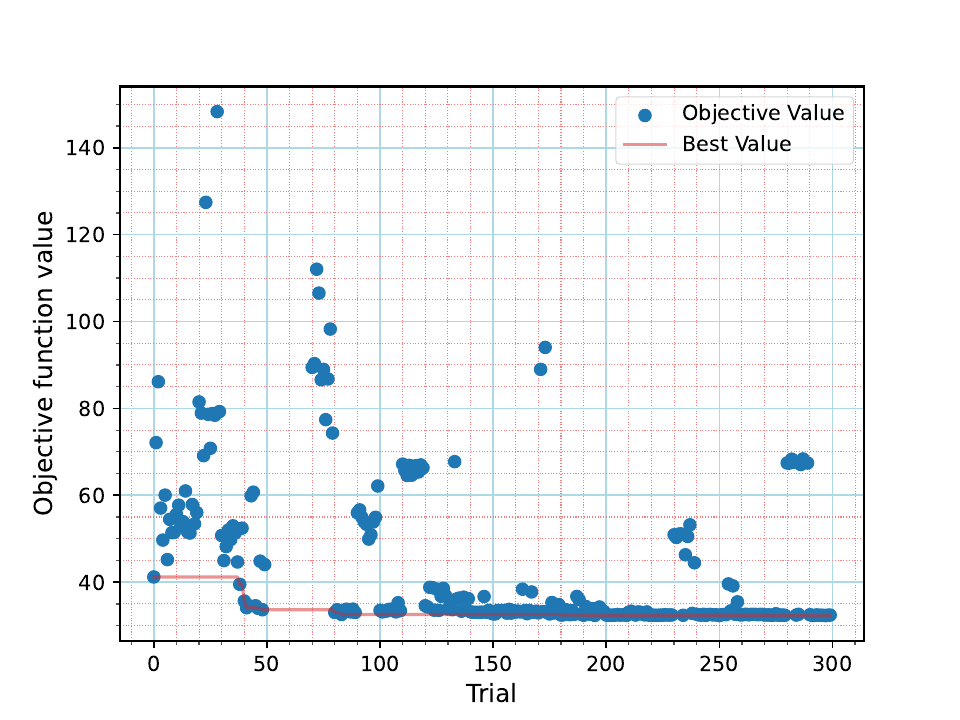}
\caption{Evolution of the objective function value across trials during the optimization process. The plot shows both the objective value for each trial (\textcolor{blue}{\textbullet}) and the best value (\textcolor{red}{-}) found up to that point, highlighting the convergence behavior of the optimization.}
\label{fig:cost_func_optim_res}
\end{figure*}

BO is implemented by coupling the TOMAWAC wave model with the open‑source optimization framework ``Optuna'' \citep{Akiba_2019} within a Python environment, using the TelApy interface of the openTELEMAC system \citep{Goeury_2022}. The optimization workflow is deployed on high‑performance computing (HPC) resources using MPI‑based parallelization, enabling concurrent execution of independent model evaluations. At each iteration, ten simulations are launched in parallel, each running on a dedicated compute node with 48 CPU cores (Intel(R) Xeon(R) Platinum 8260 processors at 2.40\,GHz, Cascade Lake architecture). A single TOMAWAC simulation requires approximately 12\,h\,30\,min of wall‑clock time, making batch‑parallel execution essential to achieve practical turnaround times and to efficiently explore the high‑dimensional parameter space within realistic computational constraints. The BO procedure is conducted under a fixed computational budget of approximately 300 trials, with minor deviations in rare cases of node failures. This fixed budget strategy accounts for the high cost of individual simulations and ensures consistency and fairness across optimization experiments. The optimization behavior and convergence are subsequently analyzed based on the evolution of the objective function and the structure of the near‑optimal solution space (Figure~\ref{fig:cost_func_optim_res}). While alternative stopping criteria based on convergence indicators, such as thresholds on expected improvement, could be considered, their practical implementation remains challenging for computationally intensive geophysical models. The investigation of adaptive termination strategies is therefore left as a perspective for future work.

As expressed in Eq.~\ref{eq:error_metrics}, the optimal estimation of the control vector $\boldsymbol{\theta}$ is formulated as the minimization of an objective function that quantifies the discrepancy between model outputs and observational data. This optimization is performed using a Bayesian approach. The proposed methodology facilitates the simultaneous calibration of the model structure, defined by the choice of constitutive law, and the associated underlying parameter, within a unified optimization framework. Based on the correlation analysis, the observation vector $\textbf{Y}^{\text{obs}}$ corresponds to the spectral significant wave height ($H_{m0}$), recorded every 30 minutes by the CANDHIS network at the following buoy locations: \textit{Plateau~du~Four}, \textit{Île~de~Sein}, \textit{Pierres~Noires}, \textit{Cap~Ferret}, \textit{Belle--Île}, and \textit{Saint--Jean--de--Luz}, over the period from February 1 to March 1, 2014. In this study, the weighting matrix $\mathcal{W}$ is specified as a diagonal matrix, with each diagonal entry representing the relative contribution of a buoy measurement. The weights are depth-dependent, favoring shallower buoys to enhance sensitivity to near-surface wave dynamics. To assess the robustness of the calibration methodology, the optimization process was repeated six times using different initial guesses. The configuration yielding the lowest objective function value was selected as the optimal solution. Its consistent retrieval in three independent trials demonstrates the robustness of the optimal value. Figure \ref{fig:cost_func_optim_res} presents the evolution of the objective function across BO trials.

The evolution of the objective function over the BO trials reveals a clear and effective calibration process. Early evaluations exhibit large misfits, with objective values often exceeding 50, indicating that many admissible parameter combinations yield poor agreement with observations. The best so far value decreases rapidly during the first 40 trials, reaching about 40 and marking an exploration dominated phase. After approximately 50 trials, improvements become more gradual, with objective values mainly ranging between 35 and 70 as the optimization progressively shifts toward exploitation and the parameter distributions sharpen. From roughly 150 trials onward, the optimization stabilizes around a minimum near 33, which is repeatedly recovered, demonstrating both convergence and robustness. Occasional higher cost evaluations persist by design, maintaining limited exploration and reducing the risk of overlooking alternative minima. Overall, the trajectory shows that BO quickly identifies and then consistently refines a well defined optimal region of the parameter space.

Analysis of the discrete model structures shows a clear convergence toward a single dominant physical configuration. Among the four tested combinations of depth‑induced breaking and wave–current interaction schemes, the pairing of the \citet{Thornton_1983} breaking formulation with the \citet{Westhuysen_2012} current induced dissipation is overwhelmingly selected, accounting for 79.7\% of all evaluations. All trials within the top 20\% of performance belong to this configuration, while alternative structures fail to deliver competitive results. Within this dominant family, the optimization consistently favors the mean frequency $f_{01}$ variant of the \citet{Thornton_1983} formulation, which appears in 65.3\% of trials and exclusively contains the highest‑performing solutions. These results demonstrate that model performance during the February 2014 calibration period is primarily controlled by structural choices, and that BO robustly identifies the configuration that best reproduces the observations.

\begin{figure}[!ht]
\centering
\includegraphics[width=0.75\textwidth]{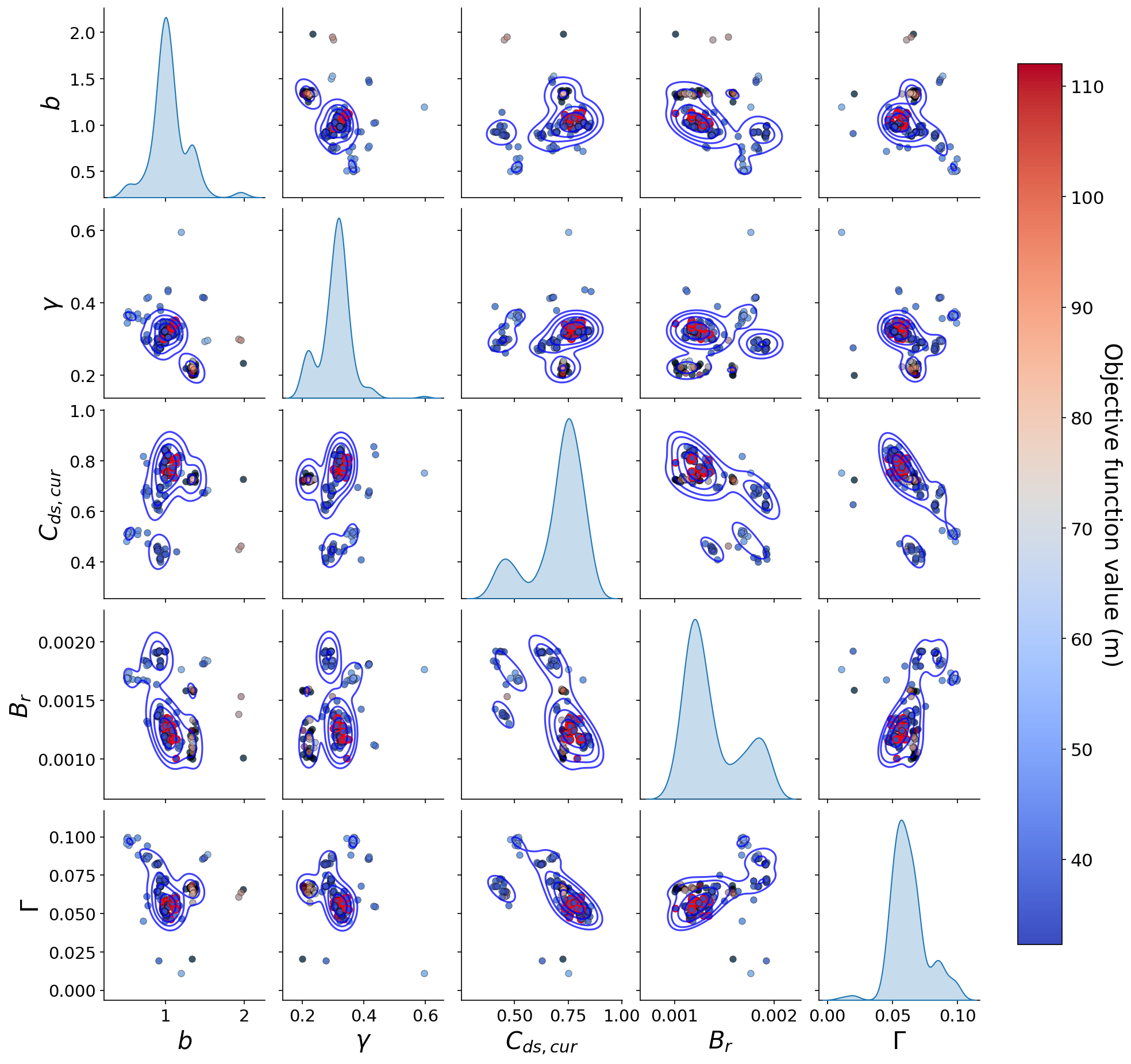}
\caption{Joint distributions of the continuous calibration parameters ($b$, $\gamma$, $C_{\mathrm{ds,cur}}$, $B_r$, and $\Gamma$) obtained from all Bayesian Optimization trials using the structural configuration that combines \citet{Thornton_1983} breaking with the representative mean frequency $f_{01}$ and the \citet{Westhuysen_2012} wave blocking by opposing current interaction formulation. Scatter points are colored by objective-function value, and red-edged markers denote the best-performing 20\% of trials. Kernel-density contours reveal the geometry of the calibration landscape, highlighting a compact, coherent near-optimal region within a broadly explored parameter domain.}
\label{fig:pairplot_parameters}
\end{figure}

To examine the influence of continuous parameters within this optimal structure, Figure~\ref{fig:pairplot_parameters} presents the joint parameter-space distributions for all BO trials restricted to the selected model configuration. Scatter points are colored by objective-function value, and the best-performing 20\% of configurations are highlighted with a red edge color, providing a clear view of the near-optimal region.

Scatter and contour plots indicate that early BO iterations broadly explore the admissible ranges of $\Gamma$, $\gamma$, $b$, $C_{\mathrm{ds,cur}}$, and $B_r$, with high‑cost evaluations distributed across all parameter combinations. In contrast, low‑error solutions (highlighted in red) cluster tightly within a single, well‑defined region, revealing a dominant basin of attraction consistent with the structural selection results. Within this region, several physically meaningful parameter interactions emerge. In the $(\gamma,b)$ plane, high‑performing solutions align along a narrow ridge defined by $\gamma \in [0.30,0.36]$ and $b \in [0.90,1.15]$, indicating a compensatory balance between breaking threshold and dissipation intensity in the \citet{Thornton_1983} formulation. The bottom‑friction coefficient converges to $\Gamma \in [0.046,0.061]$, while wave–current interaction parameters are tightly constrained to $C_{\mathrm{ds,cur}} \in [0.73,0.84]$ and $B_r \in [1.01\times10^{-3},1.40\times10^{-3}]$. No secondary low‑error clusters are observed, confirming the uniqueness and robustness of the calibrated optimum.

Overall, the joint analysis of objective function evolution, structural sampling, and continuous parameter interactions shows that Bayesian Optimization provides a balanced and comprehensive exploration of the calibration space, leading consistently to a single, physically interpretable, and operationally robust model configuration.

\subsection{Physical Performance of the Calibrated Model}
\label{subsec:calib_model}

In this part, the physical behavior of the calibrated wave model is assessed using the parameter set corresponding to the minimum of the Bayesian Optimization procedure. Specifically, we adopt the configuration corresponding to the minimum of the objective function identified during the BO search. The calibrated parameters for the Thornton and Guza model were determined to be \( b = 1.129 \) and \( \gamma = 0.344 \), while the Van der Westhuysen formulation yielded optimal values of \( C_{\text{ds,cur}} = 0.792 \) and \( B_r = 0.001 \). Additionally, the parameter \( \Gamma \), which governs wave energy dissipation due to wave-seabed interactions, was calibrated to \( \Gamma = 0.053 \). Based on this optimized parameter vector, we evaluate the model’s ability to reproduce the observed high-energy sea states and examine how the adjusted coefficients influence the representation of bottom friction, depth-induced breaking, and wave blocking by opposing currents processes across the domain. 

\begin{figure*}[!h]
\centering
\begin{tabular}{c}
\subfloat[02/01/2014-03/01/2014\label{fig:time_serie_feb_2014}]{%
\includegraphics[width=0.75\textwidth]{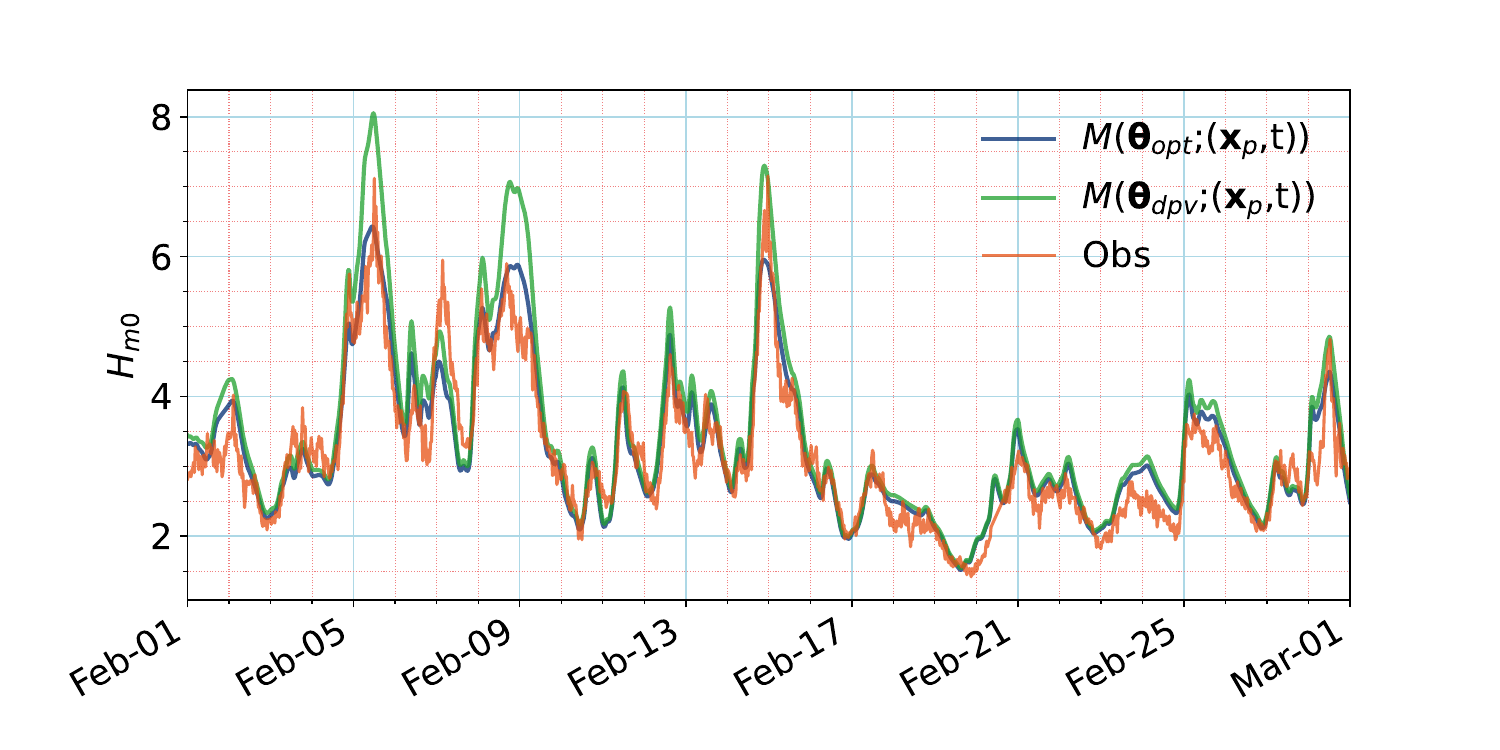}}\\
\subfloat[02/08/2014-02/18/2014\label{fig:time_serie_period}]{%
\includegraphics[width=0.75\textwidth]{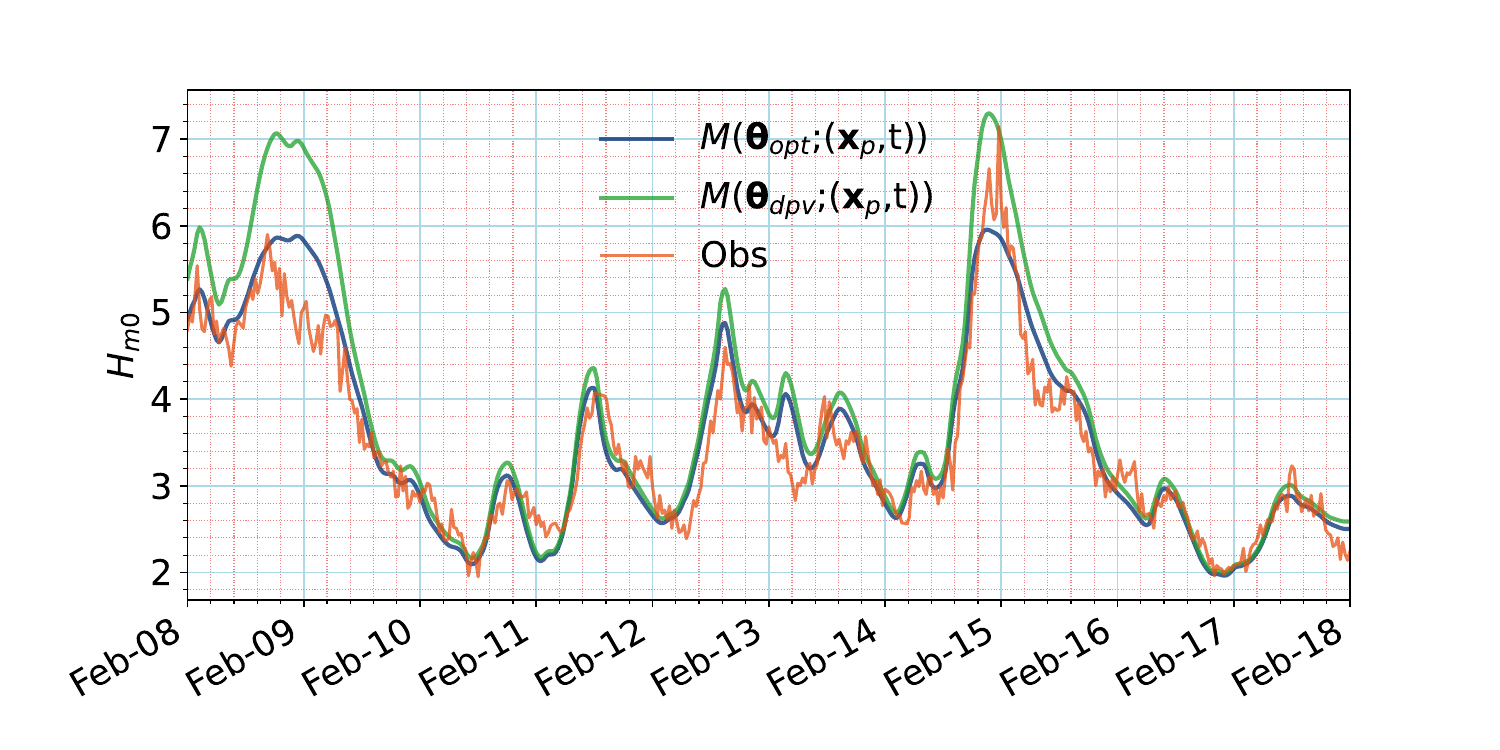}}
\end{tabular}
\caption{Comparison of the spectral significant wave height time evolution with and without calibration (respectively \textcolor{myblue}{\rule{0,5cm}{1mm}} \( M(\boldsymbol{\theta}_{opt}; ({\bf{x}}_{p}, t))\) and \textcolor{mygreen}{\rule{0,5cm}{1mm}} \( M(\boldsymbol{\theta}_{dpv}; ({\bf{x}}_{p}, t))\)) with respect to the buoy measurements at \textit{Plateau~du~Four} (\textcolor{myorange}{\rule{0,5cm}{1mm}} Obs) during February~2014.}
\label{fig:time_series_plateau_du_four}
\end{figure*}

Figure~\ref{fig:time_series_plateau_du_four} presents the time series obtained from the calibration exercise conducted at the \textit{Plateau~du~Four} buoy over the simulation period. The model output generated using the optimized parameter set, denoted
\( M(\boldsymbol{\theta}_{\mathrm{opt}}; (\mathbf{x}_p, t)) \) and hereafter referred to as the ``optimized configuration'', is compared with a reference simulation in which all model components other than the calibrated process retain their manually prescribed values, while the calibrated process uses its default parameter settings. This reference simulation, denoted
\( M(\boldsymbol{\theta}_{\mathrm{dpv}}; (\mathbf{x}_p, t)) \), is hereafter referred to as the ``default parameter configuration'' provided by the TOMAWAC solver.

\begin{figure*}[!h]
\centering
\includegraphics[width=0.75\textwidth]{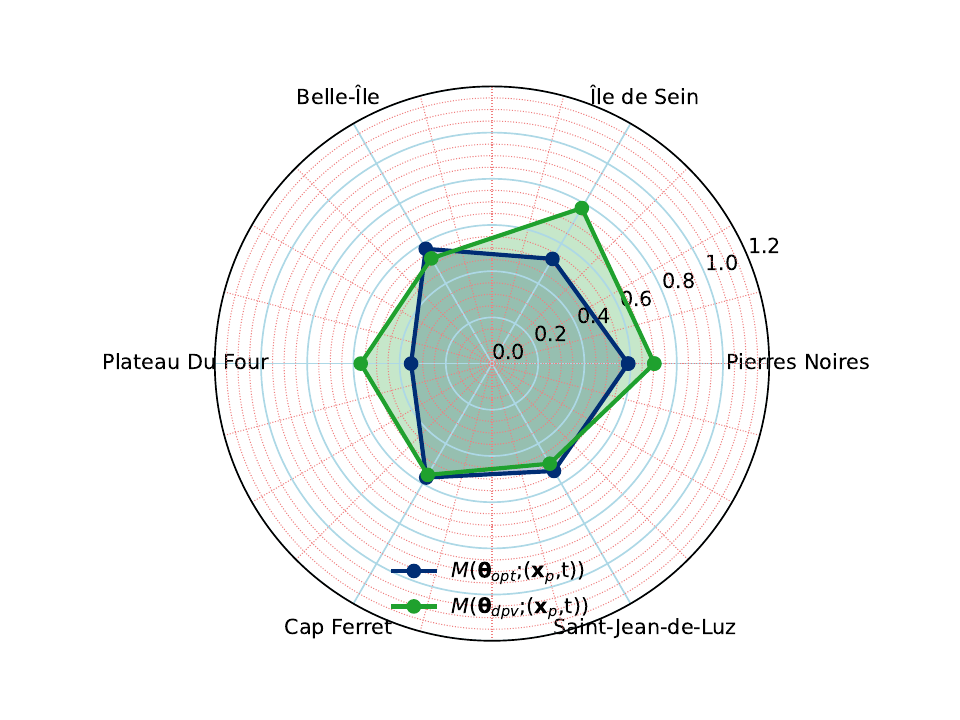}
\caption{Comparison of the root mean square error of the spectral significant wave height between the calibrated configuration \textcolor{myblue}{\rule{0,5cm}{1mm}} \( M(\boldsymbol{\theta}_{opt}; ({\bf{x}}_{p}, t))\) and the default configuration \textcolor{mygreen}{\rule{0,5cm}{1mm}} \( M(\boldsymbol{\theta}_{dpv}; ({\bf{x}}_{p}, t))\), evaluated against buoy measurements used for calibration during February 2014.}
\label{fig:spyder_cali_buoys}
\end{figure*}

As anticipated, the spectral significant wave heights computed using the calibrated parameter configuration exhibit a markedly improved agreement with observational measurements compared to those obtained using the default parameter values. Figure \ref{fig:spyder_cali_buoys} presents the Root Mean Square Error obtained for the spectral significant wave height at all buoy stations used during calibration along the French Atlantic coast. The results compare the optimized configuration, $M(\boldsymbol{\theta}_{\mathrm{opt}};(\mathbf{x}_{p},t))$, incorporating parameters obtained through BO, with the reference configuration $M(\boldsymbol{\theta}_{\mathrm{dpv}};(\mathbf{x}_{p},t))$.

RMSE values reveal strong spatial variability, with the optimized configuration consistently outperforming the default parameter reference simulation. Northern sites like \textit{Pierres~Noires} and \textit{Île~de~Sein} show lower RMSE, indicating successful calibration. In contrast, southern buoys such as \textit{Cap~Ferret} and \textit{Saint--Jean--de--Luz} display higher errors, likely due to complex local dynamics. \textit{Cap~Ferret}, for instance, is affected by energetic tides and shifting coastal morphology, which complicate wave modeling \citep{Nahon_2019}. Notably, despite their proximity, \textit{Belle--Île} and \textit{Plateau~du~Four} exhibit contrasting performance. \textit{Belle--Île} records higher RMSE values, even with calibration, likely due to its exposure to offshore swell and complex bathymetry. In contrast, \textit{Plateau~du~Four} benefits from more stable conditions, resulting in better model performance when calibrated. These results emphasize that local physical characteristics, beyond geographic proximity, play a critical role in model accuracy and calibration effectiveness.

\begin{figure*}[!h]
\centering
\begin{tabular}{cc}
\subfloat[Optimized configuration\label{fig:qq_plot_plateau_optim}]{%
\includegraphics[width=0.55\textwidth]{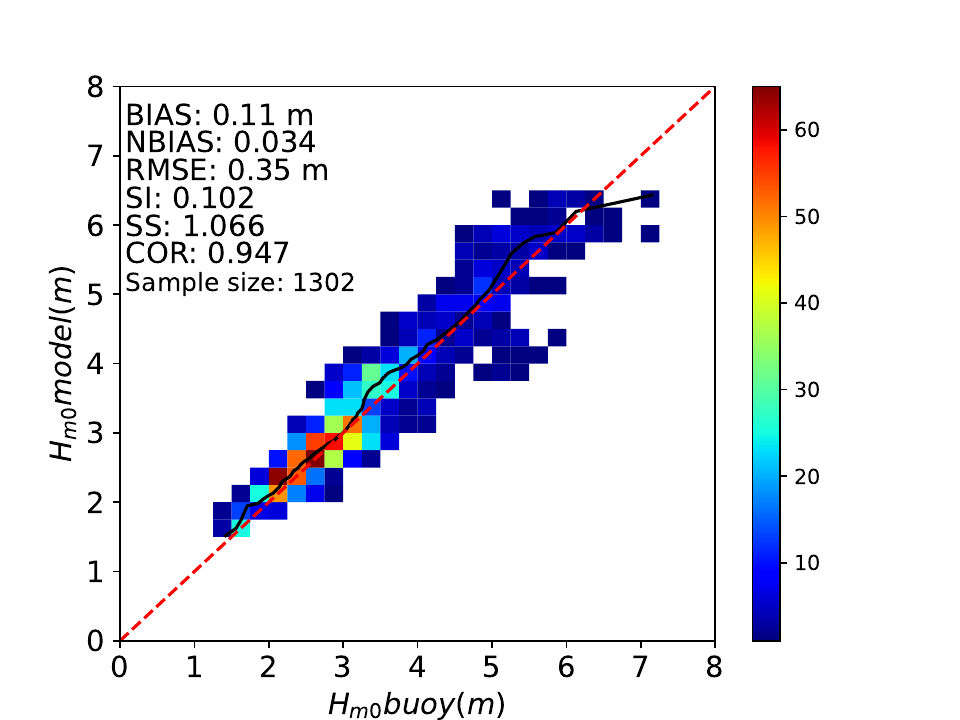}}&
\subfloat[Reference configuration\label{fig:qq_plot_plateau_pdv}]{%
\includegraphics[width=0.55\textwidth]{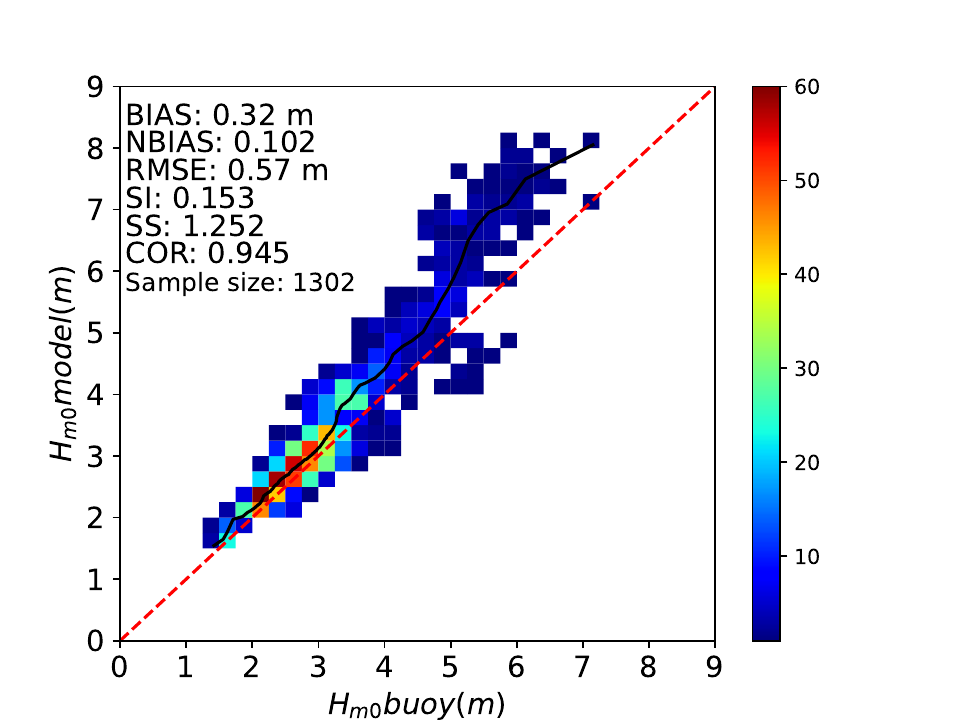}}\\
\end{tabular}
\caption{Scatter and Q–Q plots comparing modeled versus observed spectral significant wave heights ($H_{m0}$) at the \textit{Plateau~du~Four} buoy for the optimized and default parameter reference configurations. Performance metrics include bias (BIAS, representing systematic error), normalized bias (NBIAS, indicating relative error), root mean square error (RMSE, quantifying overall error magnitude), scatter index (SI, measuring dispersion), skill score (SS, assessing predictive accuracy), and correlation coefficient (COR, reflecting linear association).}
\label{fig:qq_plot_plateau}
\end{figure*}

Another key performance indicator is the scatter plot between observed versus modeled significant wave height results and the superposed the Quantile-Quantile (Q-Q) plot. A perfect agreement between the model and the observational data distributions is indicated when the Q-Q plot aligns precisely with the 45-degree diagonal line. Figure \ref{fig:qq_plot_plateau} compares modeled and observed significant wave heights ($H_{m0}$) at \textit{Plateau~du~Four} buoy for the optimized and default parameter reference configuration.

As shown in Figure~\ref{fig:qq_plot_plateau}, both models exhibit strong correlation with observations ($\mathrm{COR} \approx 0.95$), but performance metrics reveal significant differences. The optimized reanalysis configuration (Figure \ref{fig:qq_plot_plateau_optim}) demonstrates lower bias (0.11~m), RMSE (0.35~m), and scatter index (0.102), with the Q-Q plot closely following the diagonal, particularly for mid-range values (2--5~m), indicating accurate distributional representation. Minor deviations occur at higher wave heights ($>6$~m), where the model tends to underestimate extremes. Conversely, the reference reanalysis configuration (Figure \ref{fig:qq_plot_plateau_pdv}) shows higher bias (0.32~m), RMSE (0.57~m), and scatter index (0.153), with greater deviations at higher quantiles, suggesting overestimation and reduced reliability under extreme conditions. The broader dispersion observed for the default parameter reference configuration indicates greater variability and less stable performance than the optimized re‑analysis.

\newpage

\subsection{Cross-validation on independent data}
\label{subsec:cross_validation_global}

To evaluate the generalizability of the optimized parameter vector $\boldsymbol{\theta}_{opt}$ beyond the calibration configuration, two complementary cross-validation exercises were performed: (i) a spatial cross-validation using buoy stations not included in the calibration subset, and (ii) a temporal cross-validation using storm-active periods distinct from the February~2014 calibration window. Together, these tests assess whether the Bayesian Optimization (BO) procedure leads to physically meaningful and transferable parameter settings across both space and time.

\subsubsection{Spatial cross-validation on independent buoy locations}
\label{subsubsec:spatial_cross}

\begin{figure*}[!h]
\centering
\includegraphics[width=0.75\textwidth]{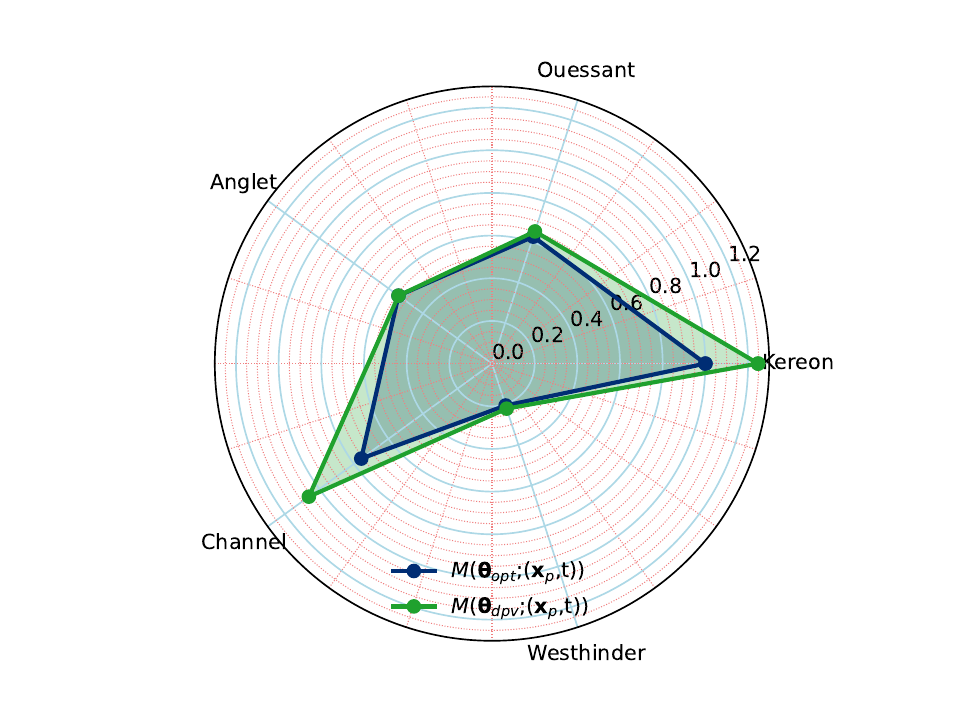}
\caption{Comparison of the RMSE of the spectral significant wave height between the calibrated configuration \textcolor{myblue}{\rule{0,5cm}{1mm}} \( M(\boldsymbol{\theta}_{opt}; ({\bf{x}}_{p}, t))\) and the default configuration \textcolor{mygreen}{\rule{0,5cm}{1mm}} \( M(\boldsymbol{\theta}_{dpv}; ({\bf{x}}_{p}, t))\), evaluated at independent validation buoys during February~2014.}
\label{fig:spyder_validated_buoys}
\end{figure*}

Figure~\ref{fig:spyder_validated_buoys} presents the RMSE computed at buoy stations deliberately excluded from the calibration process for the February~2014 period. These independent sites provide a rigorous assessment of the spatial robustness of the optimized parameter set.

Across these validation stations, the optimized reanalysis configuration generally yields improved performance relative to the reference reanalysis configuration. The most notable reductions in RMSE are observed at \textit{Kereon} and \textit{Channel}. Although \textit{Kereon} and \textit{Ouessant} are geographically close to several calibration sites, they exhibit contrasting responses to the optimization process. \textit{Ouessant}, being the deepest buoy and less influenced by the processes studied, shows limited improvement.  Conversely, the enhanced results obtained at \textit{Channel} indicate meaningful spatial transferability of the calibrated dissipation settings. At greater distance from the calibration region, the \textit{Westhinder} station exhibits more modest RMSE reductions, illustrating the difficulty of extrapolating shallow-water dissipation tuning to areas characterized by distinct bathymetric and hydrodynamic conditions.

These results demonstrate that the optimized parameter set retains much of its effectiveness at nearby or dynamically similar locations, while also clarifying the natural limits of spatial extrapolation. This highlights the importance of employing geographically distributed calibration and validation strategies when applying reanalysis methodologies over large and heterogeneous coastal domains.

\newpage

\begin{figure*}[!h]
\centering
\begin{tabular}{cc}
\subfloat[optimized configuration\label{fig:qq_plot_channel_optim}]{%
\includegraphics[width=0.55\textwidth]{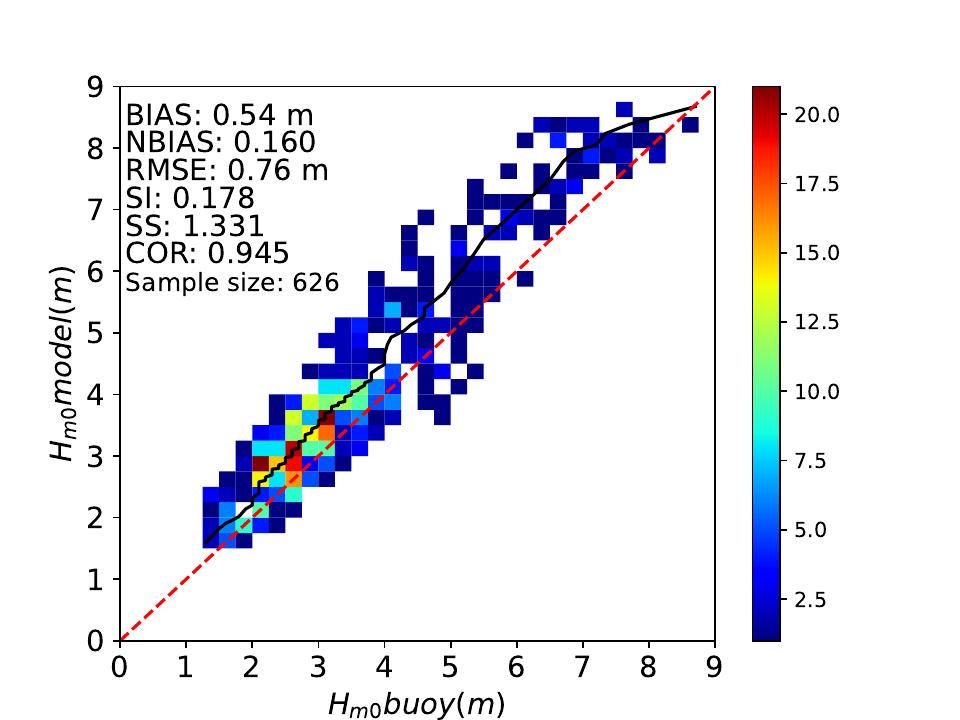}}&
\subfloat[reference configuration\label{fig:qq_plot_channel_pdv}]{%
\includegraphics[width=0.55\textwidth]{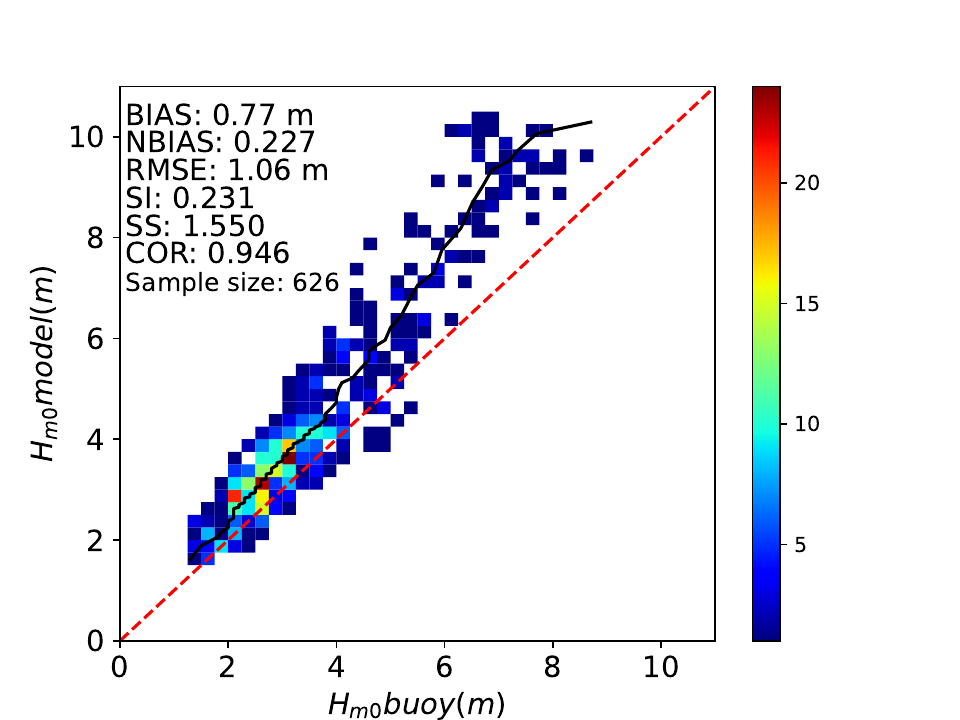}}\\
\end{tabular}
\caption{Scatter and Q–Q plots comparing modeled versus observed spectral significant wave heights ($H_{m0}$) at the \textit{Channel} buoy for the optimized and default parameter reference configurations. Performance metrics include bias (BIAS, representing systematic error), normalized bias (NBIAS, indicating relative error), root mean square error (RMSE, quantifying overall error magnitude), scatter index (SI, measuring dispersion), skill score (SS, assessing predictive accuracy), and correlation coefficient (COR, reflecting linear association).}
\label{fig:qq_plot_channel}
\end{figure*}

\begin{figure*}[!h]
\centering
\begin{tabular}{cc}
\subfloat[Optimized reanalysis\label{fig:qq_plot_ouessant_optim}]{%
\includegraphics[width=0.55\textwidth]{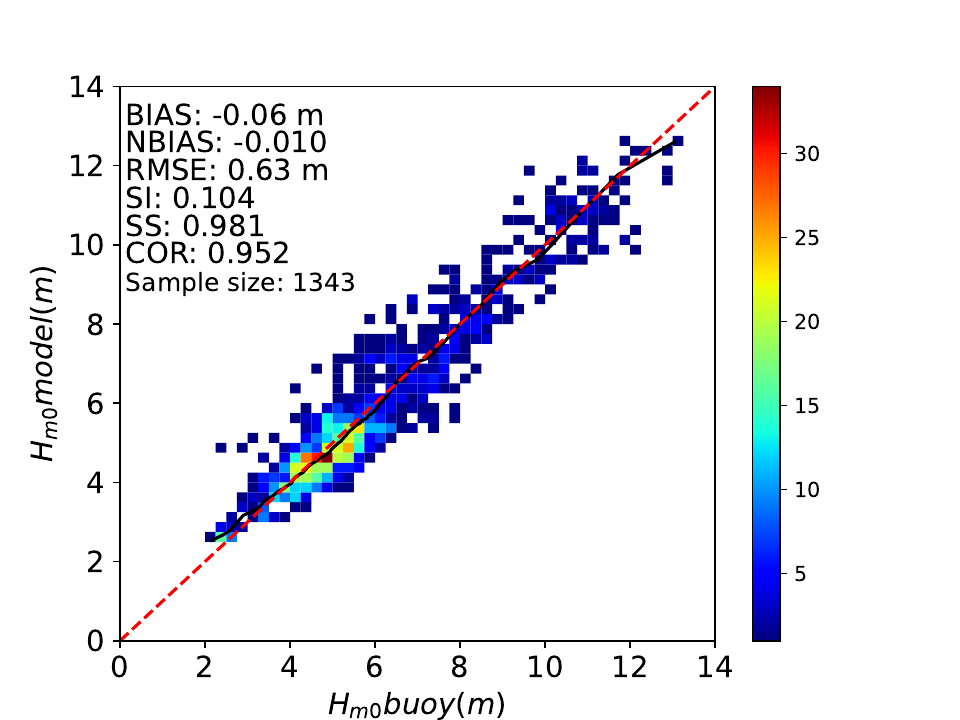}}&
\subfloat[Default reanalysis configuration\label{fig:qq_plot_ouessant_pdv}]{%
\includegraphics[width=0.55\textwidth]{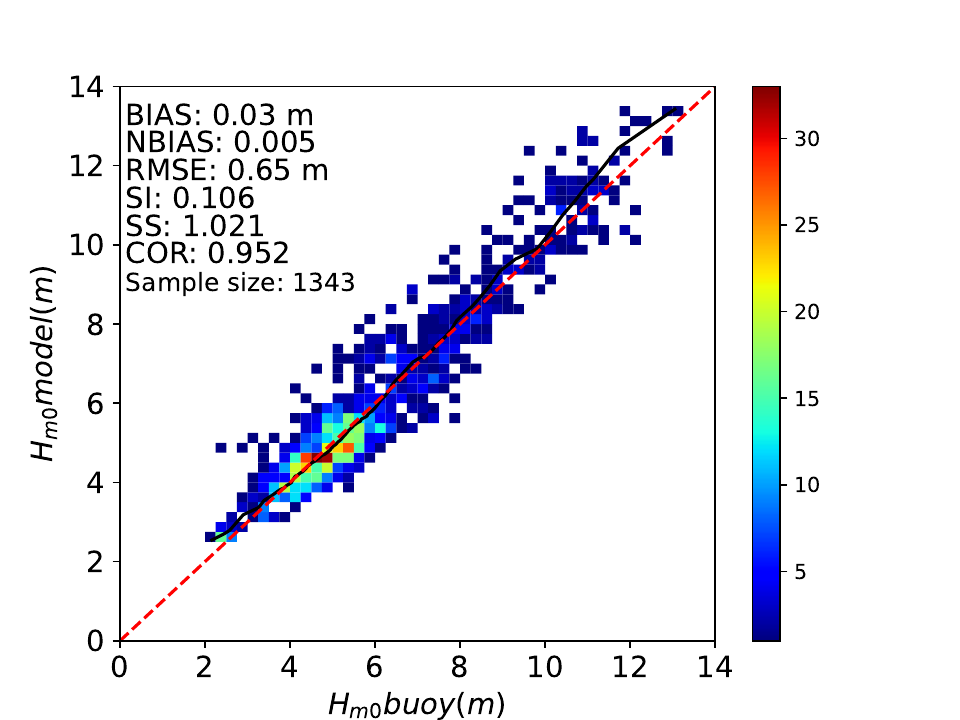}}\\
\end{tabular}
\caption{Scatter and Q–Q plots comparing modeled versus observed spectral significant wave heights ($H_{m0}$) at the \textit{Ouessant} buoy for the optimized and default parameter reference configurations. Performance metrics include bias (BIAS, representing systematic error), normalized bias (NBIAS, indicating relative error), root mean square error (RMSE, quantifying overall error magnitude), scatter index (SI, measuring dispersion), skill score (SS, assessing predictive accuracy), and correlation coefficient (COR, reflecting linear association).}
\label{fig:qq_plot_ouessant}
\end{figure*}

To complement the RMSE analysis, Figures \ref{fig:qq_plot_channel} and \ref{fig:qq_plot_ouessant} present scatter diagrams of observed versus modeled  spectral significant wave heights, together with the corresponding Quantile–Quantile (Q–Q) plots, at the \textit{Channel} and \textit{Ouessant} stations. These diagnostics provide a more detailed characterization of model behavior across the full range of observed sea states, and reveal performance differences that are not captured solely by RMSE.

Figure~\ref{fig:qq_plot_channel} illustrates the results obtained at the \textit{Channel} buoy.

As shown in Figure~\ref{fig:qq_plot_channel}, the calibrated configuration yields substantial improvements at this site. The RMSE decreases from 1.06\,m to 0.76\,m, and the positive bias is notably reduced (from 0.77\,m to 0.54\,m). The scatter cloud becomes more tightly aligned with the one-to-one line during high energy events, indicating a more accurate representation of dissipation processes during storms. The correlation remains essentially unchanged, confirming that the calibration primarily enhances amplitude accuracy rather than altering the timing of wave evolution.

Figure~\ref{fig:qq_plot_ouessant} displays the results obtained at the \textit{Ouessant} buoy.

As illustrated in Figure~\ref{fig:qq_plot_ouessant}, the influence of the  calibration at this deep-water station is minimal. The RMSE decreases only slightly (from 0.65\,m to 0.63\,m), and the correlation coefficient remains unchanged. The near-complete overlap between the calibrated and default scatter clouds confirms that the optimized parameters do not alter deep-water performance. This is consistent with expectation, as the dissipation processes targeted by the calibration have limited impact in deep-water environments. These results confirm the selective effect of the calibration, delivering substantial improvements at coastal and transition depth sites while preserving accuracy in deep water regions.

\newpage

\subsubsection{Temporal cross-validation on independent storm periods (January 2014 and 2018)}

To further assess the robustness and generalizability of the calibrated parameter set, an independent validation was carried out over two storm-active winter periods not included in the calibration process, namely January~2014 and January~2018. Both months exhibit energetic North Atlantic storm activity, yet their temporal structure and overall weather patterns differ from those observed during the February~2014 calibration window. This variability provides a meaningful basis for evaluating the ability of the optimized parameters to transfer across distinct forcing conditions and sea state evolutions.  It is noted that the observational network is not identical between the two periods, with fewer wave buoy measurements available during January~2018.

\begin{figure*}[!h]
\centering
\begin{tabular}{c}
\subfloat[January 2014\label{fig:spyder_rmse_janv_2014}]{%
\includegraphics[width=0.75\textwidth]{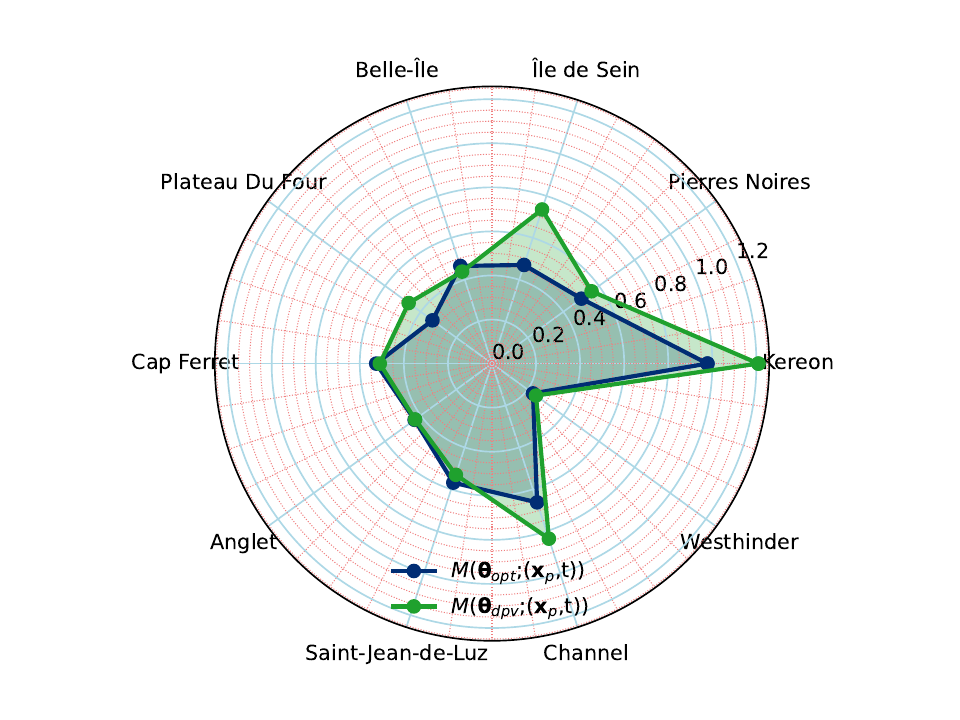}}\\
\subfloat[January 2018\label{fig:spyder_rmse_janv_2018}]{%
\includegraphics[width=0.75\textwidth]{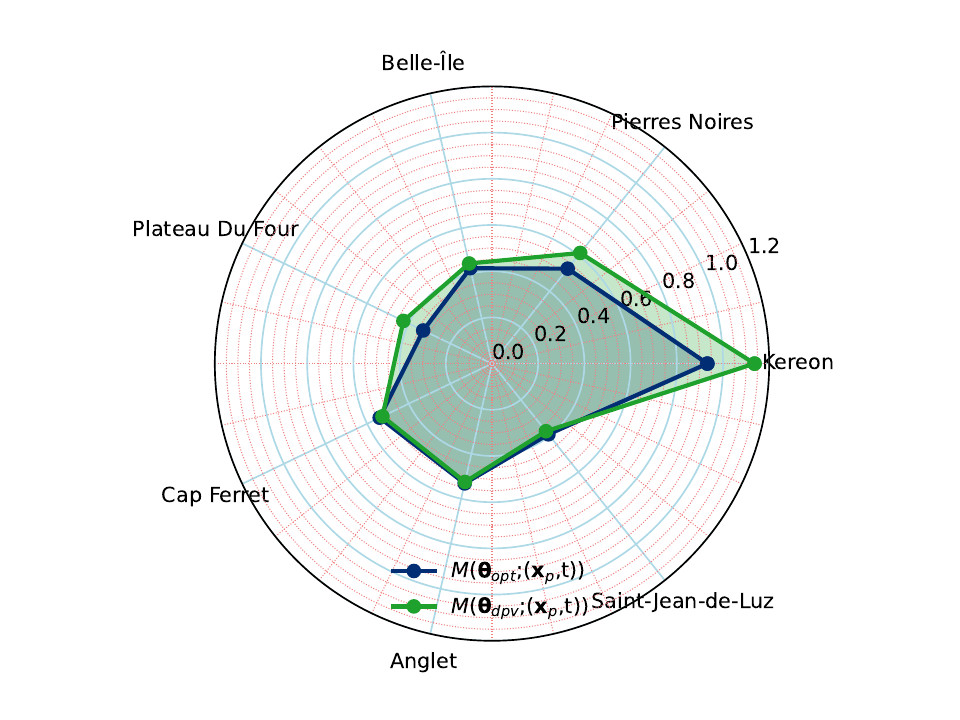}}
\end{tabular}
\caption{Comparison of the root mean square error on spectral significant wave height evolution with and without calibration (respectively \textcolor{myblue}{\rule{0,5cm}{1mm}} \( M(\boldsymbol{\theta}_{opt}; ({\bf{x}}_{p}, t))\) and \textcolor{mygreen}{\rule{0,5cm}{1mm}} \( M(\boldsymbol{\theta}_{dpv}; ({\bf{x}}_{p}, t))\) during January 2014 and 2018.}
\label{fig:spyder_rmse_janv_2014_and_2018}
\end{figure*}

Figure~\ref{fig:spyder_rmse_janv_2014_and_2018} compares the RMSE obtained with and without calibration during January~2014 (Figure~\ref{fig:spyder_rmse_janv_2014}) and January~2018 (Figure~\ref{fig:spyder_rmse_janv_2018}).

Despite differences in storm characteristics relative to the calibration period, the optimized configuration consistently outperforms the reference model at most buoy locations. Coastal stations such as \textit{Plateau~du~Four}, \textit{Pierres~Noires} and \textit{Île~de~Sein} exhibit systematic RMSE reductions, typically ranging from 15--30\%, comparable to those obtained during February~2014 (Figure~\ref{fig:spyder_cali_buoys}). \textit{Kereon} and \textit{Channel}, also exhibit robust improvements, confirming that the calibrated dissipation parameters remain effective across storm sequences with markedly different timing and intensity.

As noted in Sections~\ref{subsec:calib_model} and \ref{subsubsec:spatial_cross}, several stations exhibit only limited differences between the default and calibrated configurations. Southern buoys such as \textit{Cap~Ferret} and \textit{Saint--Jean--de--Luz} retain comparatively high RMSE values, likely reflecting the influence of complex local wave climates and coastal morphology that are only partially governed by the dissipation processes considered here. \textit{Belle--Île} also exhibits limited improvement, potentially reflecting the influence of long‑period swell and complex bathymetric gradients. \textit{Ouessant}, located in deep water, remains largely unaffected by the calibration, which is consistent with the limited influence of bottom friction and depth-induced breaking in offshore environments. Importantly, the stability of model performance at these sites indicates that the calibration does not degrade behavior outside regions dominated by dissipation processes, thereby supporting the physical consistency and selective impact of the parameter adjustments.

Taken together, the temporal cross‑validation results across three storm‑active winter periods indicate that the optimized parameter vector $\boldsymbol{\theta}_{\mathrm{opt}}$ captures robust physical behavior rather than overfitting the conditions of the calibration month. Consistent RMSE reductions at coastal and intermediate‑depth stations, together with the expected neutral response at deep‑water sites, demonstrate that the calibrated parameters generalize effectively across a broad range of energetic winter regimes. These results therefore support the suitability of the proposed optimization framework for long‑term hindcast and large‑scale reanalysis applications.

\newpage

\section{Discussion}
\label{sec:discussions}

The results obtained in this study confirm the effectiveness of BO for calibrating wave model sink terms, particularly in the context of sea state reanalysis. The integration of the TPE within a scalable optimization framework led to significant improvements in model fidelity, with systematic reductions in RMSE and bias observed across multiple buoy locations. Quantitatively, these improvements are preserved under independent spatial and temporal validation. At stations where calibration yields the most significant RMSE reductions, relative changes between calibration and validation generally remain within $\pm$10\%, while at locations where the impact of calibration is more marginal, RMSE values remain stable across periods. This behavior indicates that a substantial fraction of the calibration skill is retained outside the training dataset, despite differing storm sequences and forcing conditions. Together, these outcomes demonstrate the potential of BO to enhance the representation of complex coastal wave dynamics, especially under high-energy conditions. Nevertheless, several methodological aspects warrant further investigation to improve model performance and generalizability.

The outcomes of BO inherently depend on the formulation of the optimization problem. In particular, the definition of the error metric, the selection and weighting of observations, and the specification of the admissible parameter space jointly shape the optimization landscape explored by the algorithm. In addition, the optimization behavior is influenced by algorithmic hyperparameters, such as those controlling the exploration–exploitation trade‑off and the termination criteria. Consequently, the resulting parameter distributions reflect both the underlying physical processes and the assumptions embedded in the minimization setup. To assess the sensitivity of the results to the chosen weighting strategy, an additional optimization experiment was conducted using a uniform weighting matrix (Eq.~\ref{eq:error_metrics}), assigning equal importance to all calibration observations. The results, presented in ~\ref{sec:appendix:sens_BO}, indicate that the overall structure of the optimal solution space remains largely consistent with that obtained using depth dependent weighting. The near optimal regions remain compact and coherent, with the main differences limited to moderate adjustments in parameters associated with spatially extensive processes, most notably the bottom friction coefficient $\Gamma$ (Figure~\ref{fig:pairplot_parameters_w_sensitivity}).

Within this context, one key limitation of the current implementation is the assumption of a spatially uniform bottom friction coefficient. This simplification may overlook important spatial heterogeneities in seabed composition, sediment type, and bathymetric features, which can significantly influence wave energy dissipation. Recent work \citep{Wang_2022} has shown that applying spatially varying correction factors, defined with uniform values within subdomains to better represent bathymetric variability, can substantially improve wave transformation accuracy in coastal environments. Following this approach in future developments, using geophysical datasets or sedimentological maps to prescribe spatially varying bottom friction should allow the model to better capture local dissipation processes and enhance its performance in heterogeneous coastal environments.

Another area for improvement concerns enhancing the model's ability to represent fine-scale coastal dynamics. Although the calibration process has led to a significant reduction in discrepancies between model outputs and observations, persistent residual errors remain at specific buoy locations, such as \textit{Cap~Ferret} and \textit{Belle--Île}. These mismatches suggest that local processes, potentially influenced by complex bathymetry, tidal interactions, and nearshore dynamics, are not adequately resolved. Addressing these limitations through increased spatial resolution of the computational mesh and, where appropriate, coupling with higher-resolution hydrodynamic models could substantially improve the representation of localized wave transformations, thereby enhancing calibration accuracy and overall predictive performance.

The current optimization strategy focuses on minimizing a single objective function based on significant wave height. While effective, this approach may not fully capture the trade-offs between different aspects of model performance. Extending the calibration framework to a multi‑objective optimization setting would enable the simultaneous tuning of several performance metrics, thereby providing a more balanced representation of wave conditions. In addition, this limitation could be alleviated by optimizing the full wave spectrum rather than individual integrated parameters, as demonstrated by \citet{Alonso_2021}, who showed that spectrum based calibration strategies can capture a broader range of physical processes and improve model robustness. Incorporating such approaches in future developments would support a more holistic tuning of the model and improve its performance across a broader range of wave characteristics.

While BO efficiently identifies optimal parameter sets, it does not inherently provide a probabilistic characterization of uncertainty. A comprehensive understanding of parameter sensitivity and model robustness requires the integration of uncertainty quantification techniques. In this context, methods such as Markov Chain Monte Carlo (MCMC) sampling, particularly when combined with surrogate modeling to mitigate computational costs, offer a promising avenue for estimating posterior distributions of model parameters. It is worth noting that substantial work has already explored these probabilistic approaches in the context of wave‑model calibration and environmental modeling more broadly (e.g., studies cited in the introduction), demonstrating the relevance and maturity of Bayesian and Monte‑Carlo based frameworks for uncertainty characterization. This probabilistic insight would enable the assessment of confidence intervals associated with both model inputs and outputs, thereby enhancing the interpretability of calibration results. Such an approach is especially valuable in operational settings, where informed decision-making depends not only on model accuracy but also on a transparent evaluation of uncertainty.

Building upon this perspective, recent developments have promoted optimization under uncertainty workflows in which data-driven surrogates are trained to jointly predict central responses and associated uncertainty measures, thereby enabling explicit uncertainty propagation and reliability-informed decision-making under acceptable computational cost \citep{La_2026}. From this standpoint, the BO framework presented in this study can be regarded as a foundational and complementary step within a broader uncertainty quantification modeling chain. By efficiently constraining the admissible parameter space and identifying robust families of near-optimal solutions in a full-physics wave model, it provides a physically consistent and computationally tractable basis upon which surrogate assisted MCMC or reliability-informed optimization strategies could be naturally developed in future work.

Taken together, these avenues for refinement highlight the adaptability of the BO framework and its potential for broader application in modeling geophysical flows and waves. By addressing these methodological challenges, future work can further improve the accuracy, reliability, and utility of wave model calibration in coastal and oceanographic studies.

\section{Conclusion}
\label{sec:conclusions}

This study demonstrates the benefits of using Bayesian Optimization (BO) as a robust and efficient framework for calibrating dissipation (sink) terms in spectral wave models, with a particular focus on the ANEMOC-3 hindcast wave database. From a methodological perspective, embedding the Tree-structured Parzen Estimator (TPE) within the calibration process enables the joint optimization of continuous physical parameters and discrete model structures in a computationally demanding modeling environment. The proposed framework is scalable, fully compatible with the openTELEMAC system, and readily transferable to a wide range of geophysical modeling applications beyond wave modeling, including hydrodynamics, water quality, and morphodynamics.

From an application standpoint, the ANEMOC-3 case study shows that the calibrated configuration significantly improves model–observation agreement during high-energy sea-state conditions. The optimized setup outperforms the reference configuration across multiple buoy locations, and quantitative spatial and temporal validation demonstrates that most of these gains are preserved under independent storm periods. While the magnitude of improvement varies spatially, reflecting local dynamical regimes, the results confirm that the calibrated parameters capture robust and transferable dissipation characteristics.

Together, these findings highlight the dual contribution of this work by combining a generic and reusable optimization methodology with physically meaningful improvements obtained from a challenging real-world application. Future extensions, including multi-objective optimization and formal uncertainty quantification, represent promising avenues for future research. In summary, the use of advanced optimization strategies such as Bayesian Optimization represents a significant advancement in the calibration of complex environmental models, and this work lays the groundwork for more reliable, efficient, and adaptive modeling systems, ultimately contributing to improved ocean and coastal forecasting and hindcasting capabilities.

\section*{Acknowledgments}

This work was carried out as part of EDF R\&D's MOISE-2 research project on river and coastal flood hazard assessment, whose support the authors gratefully acknowledge. The authors gratefully acknowledge contributions from the open-source community, especially that of the open-source hyperparameter optimization framework to automate hyperparameter search library ``Optuna''. The authors also would like to thank the two anonymous reviewers, whose comments and suggestions helped improve the manuscript.


\appendix

\section{Detailed Mathematical Formulations for depth-induced breaking and wave-current interaction dissipation}

This appendix compiles the full mathematical formulations associated with the parameterizations of depth-induced breaking and wave-current interaction discussed in the main text.

\subsection{Depth-induced breaking dissipation}
\label{app:breaking}

To represent depth induced wave energy dissipation, parametric spectral models that estimate total energy loss by combining a breaking probability with a rate of energy dissipation can be adopted. Originally developed for random wave fields, these models are based on the formulations proposed by \citet{Battjes_1978} and \citet{Thornton_1983}, and are described in the following sections. Despite differences in the model implementation, all approaches share a common conceptual foundation, drawing an analogy with hydraulic jumps. In the directional spectrum formulation, it is assumed that wave breaking reduces the total energy without modifying its distribution across frequencies and directions.

The characteristic frequency \( f_c \), employed in both formulations, is chosen by the modeler. In this study, it corresponds either to the spectral peak frequency, or to representative mean frequencies such as \( f_{01} = \frac{m_1}{m_0} \), where \( m_n \) denotes the \( n \)-th order spectral moment.

\begin{itemize}
    \item Battjes and Janssen’s model
\end{itemize}

The dissipation term introduced by \citet{Battjes_1978} assumes that all breaking waves reach a characteristic height \( H_m \), which is approximately proportional to the local water depth \( h \). Based on this assumption, the breaking-induced dissipation source term \( Q_{\text{br}} \) is formulated to preserve the spectral shape of the wave energy while accounting for the total energy loss due to breaking. The dissipation is distributed proportionally across the spectrum, and the expression is given by:

\begin{equation} 
\label{eq:battjes_and_janssen} 
Q_{\text{br}} = -\frac{\alpha Q_{b} f_{c} H_{m}^{2} }{4} \cdot \frac{F}{m_{0}}
\end{equation}

where,  \( \alpha \) is a dimensionless coefficient of order 1, \( Q_b \) is the fraction of breaking waves,  \( f_c \) is a characteristic wave frequency, \( H_m \) is the maximum wave height associated with breaking, \( F \) is the directional variance spectrum, \( m_0 \) is the zeroth-order spectral moment (total wave energy).

The breaking fraction \( Q_b \) is computed from the Battjes and Janssen model as the solution of the implicit equation:

\begin{equation*}
\frac{1 - Q_b}{\ln Q_b} = -\frac{H_{m0}^2}{2 H_m^2}
\end{equation*}

where \( H_{m0} \) is the root-mean-square wave height of the incident sea state. 

The model assumes that the cumulative distribution function of wave heights is given by a Rayleigh distribution, which is abruptly truncated at $H = H_m$ to represent the physical limit imposed by wave breaking. The characteristic wave height \( H_m \) used in the breaking-induced dissipation model can be estimated either through a linear relation \( H_m = \gamma_2 h \), or using a formulation derived from Miche’s criterion:

\begin{equation*}
H_m = \frac{\gamma_1}{k_c} \tanh\left( \frac{\gamma_2 k_c h}{\gamma_1} \right)
\end{equation*}

where \( k_c \) is the magnitude of the wave number vector associated with the characteristic frequency \( f_c \), obtained from the linear dispersion relation. The empirical coefficients \( \gamma_1 \) and \( \gamma_2 \) are calibration parameters that control the limiting steepness and depth dependence of breaking.

\begin{itemize}
    \item Thornton and Guza’s model
\end{itemize}

The wave breaking model proposed by \citet{Thornton_1983} describes energy dissipation using two statistical formulations for the distribution of breaking wave heights. The first, referred to as the uniform breaking formulation in Eq.~\ref{eq:Qbr_functions}, assumes that all wave heights contribute equally to energy dissipation. The second, known as the weighted breaking formulation (also in Eq.~\ref{eq:Qbr_functions}), gives greater weight to waves with larger heights, reflecting observational evidence that breaking predominantly affects higher waves. This refinement yields a more realistic representation of wave breaking in natural sea states. The corresponding energy sink term \( Q_{\text{br}} \) is expressed according to the selected wave height distribution.

\begin{equation}
Q_{br} = 
\left\{
\begin{array}{ll}
\text{Uniform breaking :} & -48\sqrt{\pi} b^3 f_c \dfrac{(2m_0)^{5/2}}{H_m^4 h} \; F \\
\text{Weighted breaking:} & -12\sqrt{\pi} b^3 f_c \dfrac{(2m_0)^{3/2}}{H_m^2 h} \left[1 - \left(1 + \dfrac{8m_0}{H_m^2} \right)^{-5/2} \right] \; F
\end{array}
\right.
\label{eq:Qbr_functions}
\end{equation}

Here, \( f_c \) denotes the characteristic wave frequency. The parameter \( b \), typically ranging from 0.8 to 1.5, serves as a calibration coefficient. The maximum wave height compatible with the local water depth, \( H_m \), is defined as \( H_m = \gamma h \), where \( \gamma \) is an empirical breaking parameter.

\subsection{Wave blocking effects}
\label{app:blocking}

Wave blocking occurs when surface waves propagate against a strong opposing current and the current velocity approaches the wave group velocity, leading to a reduction or cessation of wave energy transmission. Accurate representation of this process is critical for realistic wave modeling in regions with strong ambient flows. This study evaluates two common approaches for incorporating wave blocking in spectral wave models: equilibrium spectrum limitation, which constrains wave energy based on physical thresholds, and dissipative source terms, which simulate energy loss due to blocking.

\begin{itemize}
    \item Equilibrium spectrum limitation
\end{itemize}

This method applies an upper bound to the wave energy spectrum based on a Phillips-type formulation \citep{Phillips_1977}. When the spectral energy exceeds a threshold defined by a \( f^{-5} \) decay law, the spectrum is rescaled to maintain consistency with the theoretical maximum energy, given by Phillips’s constant (\( \alpha_p = 0.0081 \)):

\begin{equation}
    \text{If } E(f) > E_{\text{max}} = \frac{\alpha_p g^2}{(2\pi)^4 f^5}, \quad \text{then } F = \frac{E_{\text{max}}}{E} F
\end{equation}

This approach ensures that the high-frequency tail of the spectrum remains physically realistic under strong opposing currents.

\begin{itemize}
    \item Dissipative source terms for wave-current interaction ($Q_{ds,cur}$)
\end{itemize}

Wave blocking is represented through an enhanced dissipation mechanism proposed by \citet{Westhuysen_2012}, which increases whitecapping dissipation in regions where wave steepness is amplified by opposing currents. The additional source term is defined as:

\begin{equation}
    Q_{\text{ds,cur}} = -C_{\text{ds,cur}} \max\left( \frac{\dot{f_r}}{f}, 0 \right) \left( \frac{B(k)}{B_r} \right)^{p_0 / 2} F
\end{equation}

Here, \( \dot{f_r} \) is the intrinsic frequency shift due to the current. The saturation parameter is given by \( B(k) = C_g k^3 \frac{E(f)}{2\pi} \), with \( C_g \) denoting the group velocity. The exponent \( p_0 \) is defined as:

\begin{equation}
    p_0 = 3 + \tanh\left[ w\left( \frac{u_*}{C} - 0.1 \right) \right]
\end{equation}

where \( w = 25 \) as used in \citet{Westhuysen_2007}, \( u_* \) is the friction velocity, and \( C \) is the phase speed. The parameters \( B_r \) and \( C_{\text{ds,cur}} \) are model constants representing the saturation threshold and dissipation scaling, respectively.

\section{Sensitivity of the Bayesian Optimization to Observational Weighting}
\label{sec:appendix:sens_BO}

\begin{figure}[!h]
\centering
\includegraphics[width=0.75\textwidth]{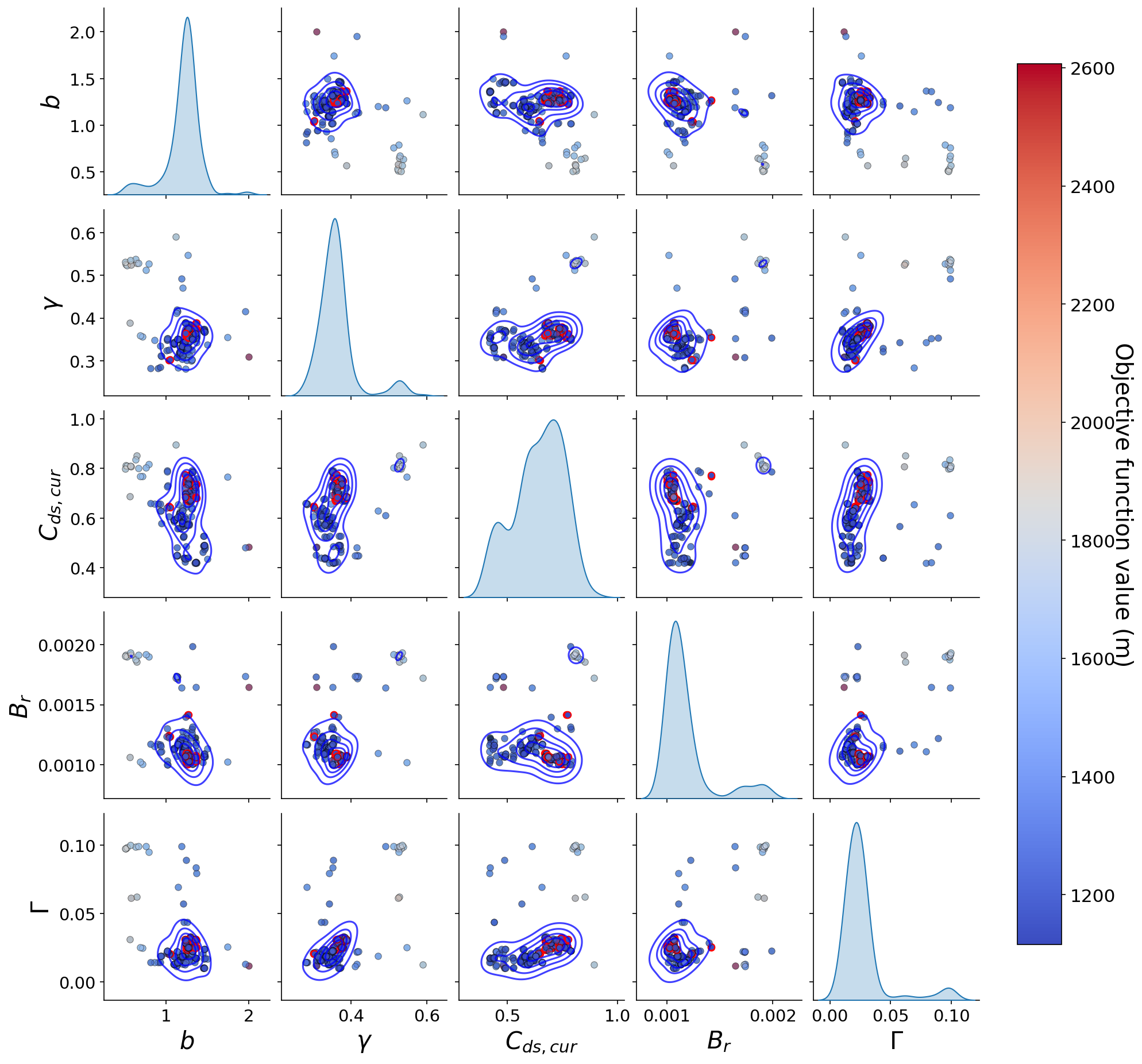}
\caption{Joint distributions of the continuous calibration parameters ($b$, $\gamma$, $C_{\mathrm{ds,cur}}$, $B_r$, and $\Gamma$) obtained from all Bayesian Optimization trials using the structural configuration that combines \citet{Thornton_1983} breaking with the representative mean frequency $f_{01}$ and the \citet{Westhuysen_2012} wave-current interaction formulation. Scatter points are colored by objective-function value, and red-edged markers denote the best-performing 20\% of trials. Kernel-density contours reveal the geometry of the calibration landscape, highlighting a compact, coherent near-optimal region within a broadly explored parameter domain.}
\label{fig:pairplot_parameters_w_sensitivity}
\end{figure}

This appendix presents a detailed analysis of the parameter-space structure
obtained from a sensitivity experiment using a uniform weighting matrix in
the error metric (Eq.~\ref{eq:error_metrics}). The experiment confirms that the structural model selected by the Bayesian Optimization remains invariant across weighting strategies. Regardless of the choice of weighting matrix in the error metric, the optimization systematically converges towards the structural configuration that combines \citet{Thornton_1983} breaking with the representative mean frequency $f_{01}$ and the \citet{Westhuysen_2012} wave-current interaction formulation. This configuration accounts for 205 trials, representing 68.3\% of all evaluated solutions, and includes all of the best-performing 20\% of optimization outcomes. Within this dominant structural configuration, the associated continuous parameters ($b$, $\gamma$, $C_{\mathrm{ds,cur}}$, $B_r$, and $\Gamma$) exhibit compact and well-defined near-optimal regions, as shown in
Figure~\ref{fig:pairplot_parameters_w_sensitivity}.

Across weighting strategies, the optimization consistently identifies a compact and coherent region of near‑optimal solutions within a broadly explored parameter space. The most pronounced sensitivity is observed for the bottom friction coefficient $\Gamma$, reflecting its role as a spatially extensive dissipation mechanism. In contrast, parameters associated with depth‑limited or locally activated processes remain comparatively stable. This behavior indicates that the weighting strategy influences the calibration primarily at the parameter level, without altering the structural model selection or the physical hierarchy of dissipation mechanisms identified by the optimization.



\end{document}